\documentclass[12pt]{article}
\usepackage{amsfonts,amsmath,amssymb,bm,graphicx,natbib,overpic}
\usepackage[dvipdfmx]{hyperref}
\bibpunct[,]{[}{]}{,}{n}{}{,}

\newcommand{\nn}{\nonumber\\}
\makeatletter

\@addtoreset{equation}{section}
\makeatother

\begin{document}
\begin{titlepage}
\pagestyle{empty}
\begin{flushright}
KEK-TH-1449
\end{flushright}

\bigskip

\begin{center}
\noindent{\large \textbf{Stochastic Equations in Black Hole Backgrounds \\
and Non-equilibrium Fluctuation Theorems}}\\
\vspace{2cm} 
\noindent{Satoshi Iso \footnote{E-mail address: satoshi.iso@kek.jp} and 
Susumu Okazawa \footnote{E-mail address: okazawas@post.kek.jp}}\\
\vspace{1cm}
  {\it
KEK Theory Center, Institute of Particle and Nuclear Studies,\\
High Energy Accelerator Research Organization(KEK)\\
and\\
The Graduate University for Advanced Studies (SOKENDAI),\\
Oho 1-1, Tsukuba, Ibaraki 305-0801, Japan
}
\end{center}


\vspace{1cm}
\begin{abstract}
We apply the  non-equilibrium fluctuation theorems 
developed in the statistical physics to the thermodynamics of black hole
horizons. In particular, we consider a scalar field in a black hole background.
The system of the scalar field behaves stochastically
due to the absorption of energy into the black hole and emission of the
Hawking radiation from the black hole horizon. 
We derive the stochastic equations, i.e. Langevin 
and  Fokker-Planck equations for a  scalar field 
in a black hole background in the $\hbar \rightarrow 0$ limit with
the Hawking temperature $\hbar \kappa/2 \pi$ fixed.
We consider two cases,  
one confined in a box with a black hole at the center 
and the other in contact with a heat bath with temperature
different from the Hawking temperature.
In the first case, the system eventually becomes  equilibrium  
with the Hawking temperature while in the second case there is an
energy flow between the black hole and the heat bath.
Applying the fluctuation theorems to these cases, we derive
the generalized second law of black hole thermodynamics.
In the present paper, we treat the black hole as a constant background
geometry.

Since the paper is also aimed to connect two different areas
of physics, non-equilibrium physics and black holes physics, 
we include  pedagogical reviews on the stochastic approaches to
the non-equilibrium fluctuation theorems and some basics of
black holes physics.
\end{abstract}
\end{titlepage}
\tableofcontents
\newpage

\section{Introduction}
The analogy of the space-time with horizons 
and thermodynamic systems have been extensively investigated,
especially, in the black hole thermodynamics \cite{Bardeen:1973gs}. 
A black hole behaves like a blackbody with the Hawking temperature
$T_{H}=\hbar \kappa/2\pi$ \cite{Hawking:1974sw}, and energy flowing into
the black hole can be identified as the entropy increase
of the black hole. 
Here, $\kappa$ is the surface gravity
at the horizon and the entropy of the black hole
$S_{BH}$ is proportional to the area of the horizon $A$ as
$S_{BH} = A/4G$ in the Einstein-Hilbert theory of gravity.
The thermal behavior is essentially quantum mechanical.
Furthermore, such a thermal behavior is not restricted to
globally-defined horizons like an event horizon of a black
hole, but also applicable to local horizons such as the
Rindler horizon of a uniformly accelerated observer. 
At the quantum level, the notion
of horizon entropy must have more fundamental meanings, since
it gives transition rates of area-changing irreversible processes
of black holes, and will be related to the quantum
statistical nature of the space-time. 
Such microscopic views have been proposed in string theory,
especially in the approach based on D-brane constructions \cite{Strominger:1996sh}.

Because of the thermodynamic behavior of black holes, a system 
in a black hole background behaves as a system in contact with a thermal
bath. In particular, if we consider a (scalar) field in a black hole background,
its effective equation must be described by a stochastic equation 
with dissipation and quantum noise.
The dissipation comes from the classical causal property of the horizon;
the black hole horizon absorbs matter and, once they fall in, they 
cannot come out. 
The property is the basis of the membrane paradigm of the black hole \cite{Thorne:1986},
in which Ohm's law or the Navior Stokes equations hold on the membrane at the
(stretched) horizon.
On the other hand, the noise term (or fluctuation) comes from the Hawking 
radiation, which is essentially quantum mechanical
and, hence, we need to quantize the system in the black hole background
in an appropriate way. The first purpose of the present paper is to derive 
such a stochastic equation of motion for a scalar field in a black hole background.
The stochastic equation of motion of a string 
is previously derived in \cite{deBoer:2008gu, Atmaja:2010uu} based on physical
intuition of the Hawking radiation, or in \cite{Son:2009vu} by using an analogy with the 
Schwinger-Keldysh formalism in the context of AdS/CFT correspondence\cite{Herzog:2002pc}.
Our approach is similar to them, but 
we obtain the effective equation by explicitly integrating fluctuating degrees of 
freedom. Namely, we introduce infinitely many variables between the horizon 
and the stretched horizon and consider them as environmental variables.
By integrating them, we can show that the variable at the stretched horizon 
behaves stochastically with a noise term. 
Though the environmental variables are living
 outside of the horizon, they 
can encode information in the black hole through choosing the Kruskal vacuum
with the regularity condition at the horizon.
In this sense, the integration of the environmental variables corresponds
to integrating hidden variables in the horizon.
The derivation of the Langevin equation is one of our main results.

The second purpose of the paper is to apply the non-equilibrium fluctuation theorem 
\cite{Evans:1993}-\cite{Jarzynski:1997} developed in 
the statistical physics to the scalar field in the black hole background.
In thermodynamic systems, entropy is always increasing (or remaining a constant).
But for a mesoscopic system where fluctuations are large, there are nonzero
probabilities that the entropy of the system decreases.
The fluctuation theorem
relates probabilities of  entropy decreasing processes to those of entropy increasing ones
in terms of the equilibrium thermodynamic quantities.
It is a very general theorem that can hold
for various dynamical and non-equilibrium systems including classical Hamilton dynamics
in contact with a heat bath, stochastic equations with dissipation and noise,
or quantum mechanical systems.
The Jarzynski equality can be derived from  the fluctuation theorem,
and the second law of thermodynamics is implied from the Jarzynski equality.
We use the word {\it implied} here because the second law can be derived only 
if we assume that a system is relaxed to an equilibrium state after a long time.
An application of the fluctuation theorem to a scalar field in a black hole
background is straightforward  once we obtain a stochastic equation of motion.
We can derive the generalized second law of black hole
thermodynamics, or Green-Kubo formula of the linear response and its nonlinear
generalizations.

The paper is organized as follows.
In section \ref{sec-SEM}, we briefly review the stochastic approach to 
thermodynamic systems, Langevin equation and Fokker-Planck equation.
An important property of the stochastic equation is that it
violates the time reversal symmetry which can be measured by an
entropy increase in the path integral.
In the next section \ref{sec-FT}, the fluctuation theorem 
for a stochastic system is reviewed. 
It relates the entropy increasing and decreasing probabilities.
From the fluctuation theorem, the Jarzynski equality is derived.
In section \ref{sec-BH}, we derive an effective stochastic equation
of a scalar field in a black hole background.
In deriving the Langevin equation, the quantum property
of the vacuum with a regularity condition at the horizon 
is very important, which is first explained. We then introduce
a set of discretized equations of a scalar field near the 
black hole horizon, and integrate the variables between the horizon
 and the stretched horizon. The integration leads to an effective 
stochastic equation for a variable at the stretched horizon.
This has the same spirit as  deriving a Langevin equation 
of a system in contact  with a thermal bath \cite{Feynman:1963, Caldeira:1981rx, Caldeira:1982uj}.
In section \ref{sec-FTBH}, we apply the fluctuation theorem 
 to the scalar field in a black hole background.
We consider two different situations. In the first case, 
we put the scalar field and the black hole in a box
with an insulating wall.
By applying the fluctuation theorem, we can derive a relation
connecting entropy decreasing probabilities with increasing ones.
The ratio is given by the difference of free energies.
From this, the generalized second law of black hole thermodynamics
can be derived.
In the second case, the wall 
is assumed to be 
in contact with a thermal bath of a different temperature
which is slightly lower than 
the Hawking temperature of the black hole.
Then there is an energy flow from the black hole to the wall.
By applying the fluctuation theorem to it, a linear response
theorem of an energy flow to the temperature difference can be
obtained.
In the appendix \ref{app-OM}, we review a derivation of the path integral
form of the Fokker-Planck equation.
In the appendix \ref{app-HR},
we will discuss the relation between 
 the noise correlation  and the flux of the Hawking radiation.
In the appendix \ref{A-SSFT}, we explain the fluctuation theorem for a steady state
and derivations of nonlinear generalizations of Green-Kubo formula.

\section{Stochastic Equations of Motion \label{sec-SEM}}
We first briefly review stochastic approaches
to classical statistical systems. In particular, we focus on 
the path-integral representation (Onsager-Machlup formalism)
of the Fokker-Planck equation and emphasize the 
role of time-reversal symmetry.
Readers familiar with non-equilibrium statistical physics 
can skip this and the next sections.

\subsection{The Langevin Equation}
The Langevin equation is a phenomenological equation of motion of a particle
with a friction term and thermal noise.
It is commonly described as 
\begin{align}
m\dot{v}&=-\gamma v -\frac{\partial V}{\partial x}+ \xi.
\label{Langevin}
\end{align} 
$V(x)$ is an external potential for the particle.
 $\gamma$ is the friction coefficient
and  $\xi(t)$ is a thermal noise (or a random force)
which is often assumed to have a Gaussian 
and white-noise (delta-correlated) distribution
\begin{align}
\langle \xi(t)\rangle=0 \ , \ \ \langle \xi(t)\xi(t')\rangle=2\gamma T \delta(t-t'). 
\end{align} 
The coefficient $2\gamma T$ is determined to satisfy the equipartition theorem
with the temperature $T$
through the fluctuation-dissipation theorem.
The noise average $\langle \cdots \rangle$ can be represented by
the following path integral  
\begin{align}
\langle F(t) \rangle&= \int {\cal D}\xi F(t) \exp\left[-\frac{1}{2}\int dt_1dt_2 \xi(t_1) \frac{\delta(t_1-t_2)}{2\gamma T} \xi(t_2)\right]
\label{NPI}
\end{align} 
with a normalization condition $\langle 1 \rangle =1$.
If necessary, we can easily generalize the noise correlation 
to an arbitrary colored non-Gaussian noise.
An well-known example that can be conveniently described by the 
Langevin equation is
the Brownian motion of a particle or thermal fluctuations of 
an electric circuit voltage.

\subsection{The Fokker-Planck Equation}
From the Langevin equation, we can derive another type of 
a stochastic equation, the Fokker-Planck equation.
It describes a dynamical evolution of the probability distribution 
$P(X,t)$ of observables $X$ at time $t$.
Here $X$ represents the variables $(x, v=\dot{x})$.
If the process is Markovian, i.e. the next state is  determined only 
by the present state, 
the time evolution of $P$ is given by the following Master equation, 
\begin{align}
\partial_t P(X,t|X_0, 0)=\int dX' \left[
w(X'\to X) P(X',t|X_0, 0) - w(X\to X')P(X,t|X_0, 0)\right] .
\label{Master}
\end{align}
Here $P(X,t | X_0,0) $ is a conditional probability to 
find an event $X(t)=X$ 
that has started from the initial value $X(0)=X_0$ at $t=0$,
i.e. $P(X,t=0|X_0,0)=\delta(X-X_0)$.
$w(X'\to X)$ is a transition rate from $X'$ to $X$,
which can be related to the Langevin equation in the following way.
The first and the second terms of the right hand side of \text{eq}.(\ref{Master}) describe 
an incoming and outgoing fluxes of $X$ respectively.

The Master equation can be brought into the Kramers-Moyal form as
\begin{align}
&\partial_t P(X,t|X_0, 0)\nn
&=-\int dr \left[w(X\to X+r)P(X,t|X_0, 0)
-w(X-r\to X) P(X-r,t|X_0, 0) \right] \nn
&=-\int dr \left[1-e^{-r\partial_X} \right] w(X\to X+r)P(X,t|X_0, 0)\nn
&=\sum_{n=1}^\infty \frac{(-1)^n}{n!}\partial_X^n \left[C_n(X) P(X,t|X_0, 0)\right],
\end{align}
where we have defined
\begin{align}
C_n(X)= \int dr r^n w(X\to X+r)= \lim_{\Delta t\to 0}
\frac{1}{\Delta t}\langle(X(t+\Delta t)-X(t))^n \rangle|_{X(t)=X}.
\end{align}
In the last line, we have rewritten  the $n$-th moment of the transition rate 
by a thermal average of an 
infinitely small variation of the observable $X$.
In this way, we can convert the Langevin equation for  dynamical variables
to the Fokker-Planck equation for the distribution functions.
Here  we show 
an explicit derivation of the Fokker-Planck equation 
for the simplest Langevin equation 
(\ref{Langevin}) as a demonstration.
Eq.(\ref{Langevin}) can be considered as a set of first order
differential equations for two variables  $x
$ and $v=\dot{x}$.
Then the Kramers-Moyal coefficients up to the second moments are
given by 
\begin{align}
C_1(x)&=  v  \nn 
C_1(v)&= -\frac{\gamma}{m}v-\frac{1}{m}\frac{\partial V}{\partial x} \nn 
C_2(x)&=0\nn
C_2(v)&=\lim_{\Delta t\to 0}\frac{1}{\Delta t}
\int_t^{t+\Delta t} dt_1 \int_t^{t+\Delta t} dt_2\langle\dot{v}(t_1)\dot{v}(t_2) \rangle|_{x(t)=x}\nn
&=\lim_{\Delta t\to 0}\left( \frac{1}{\Delta t}\int_t^{t+\Delta t} dt_1\frac{2\gamma T}{m^2}+{\cal O}(\Delta t)\right)\nn
&=\frac{2\gamma T}{m^2}.
\end{align}
Higher order coefficients vanish 
in the $\Delta t \rightarrow 0$ limit.
Now we get the Fokker-Planck equation corresponding to
the Langevin equation (\ref{Langevin});
\begin{align}
\partial_t P(x,v ,t|x_0,v_0, 0)&=\partial_x\left(-v P\right)
 +\partial_v\left[\left(\frac{\gamma}{m}v+\frac{1}{m}\frac{\partial V}{\partial x} \right)P \right]
+\partial_v^2\left(\frac{\gamma T}{m^2}P\right) .
\label{FP}
\end{align}
This Fokker-Planck equation has a simple solution 
\begin{align}
P^{\text{st}}\propto e^{-\frac{1}{T}\left(\frac{1}{2}mv^2 +V(x)\right)}.
\label{Boltzmann}
\end{align}
Note that both of 
$-v \partial_x P +\frac{1}{m}\frac{\partial V}{\partial x} \partial_v P$ 
and $\partial_v\left[\frac{\gamma}{m}v P+\frac{\gamma T}{m^2}\partial_v P\right]$ 
cancel for $P^{\text{st}}$.
It is the well-known Maxwell-Boltzmann distribution for
a system in an equilibrium with temperature $T$, and
satisfies the stationarity condition $\partial_t P^{\text{st}}=0$.
The solution satisfies the equilibrium condition, stronger than 
the stationarity condition.

Here we have used the words  "stationary" and "equilibrium" in the following sense.
Stationary distributions are solutions to
 the Fokker-Planck equation satisfying  $\partial_t P=0$.
Equilibrium distributions are also stationary but 
satisfy a stronger condition 
which is called the detailed balance condition.
The most direct definition of the detailed balance condition 
is given in the language of  the Master equation.
Due to the definition of stationarity, $P^{\text{st}}$ satisfies $\int dX' \left[w(X'\to X) P^{\text{st}}(X') - w(X\to X')P^{\text{st}}(X)\right] =0 $
for arbitrary $X$.
On the other hand, the detailed balance condition is defined as
\begin{align}
\forall X, X' , \ \ w(X'\to X) P^{\text{st}}(X') - w(X\to X')P^{\text{st}}(X)=0 .
\end{align}
To satisfy this condition, the system must have the
microscopic time reversal symmetry and
 can not have a specific arrow of time.
In other words, there is no entropy production.
In a stationary but non-equilibrium configuration, 
there is a flow of current in a configuration space $(x,v)$.

The solution of the Fokker-Planck equation can be 
represented in a path integral form as
\begin{align}
P(x,t|x_0,0)=\int_{x(0)=x_0}^{x(t)=x}{\cal D}x \exp \left[-\tfrac{1}{4\gamma T}\int_0^t dt'
\left(m\ddot{x}+\gamma \dot{x} +\tfrac{\partial V}{\partial x}\right)^2\right]
\label{Path}
\end{align}
Its derivation is explained in the appendix \ref{app-OM}.
The "Lagrangian" $L=\tfrac{1}{4\gamma T} (m\ddot{x}+\gamma \dot{x} +\tfrac{\partial V}{\partial x} )^2$
is called the Onsager-Machlup function \cite{Onsager:1953}. 
A variation of the Onsager-Machlup function gives the most probable path in the stochastic processes.
Apparently, since we have $L\geq 0$, 
the paths  satisfying $L=0$ are most favored if exist.

The Onsager-Machlup function can be divided into two parts,
\begin{align}
\frac{1}{4\gamma T} \left(m\ddot{x} +\tfrac{\partial V}{\partial x}\right)^2
+\frac{\gamma}{4T}\dot{x}^2
\end{align}
which preserves time reversal symmetry,
and a violating term,
\begin{align}
 -\frac{1}{2T}\dot{x}\left(m\ddot{x} +
\tfrac{\partial V}{\partial x}\right).
\end{align}
The latter plays an important role to prove the
fluctuation theorem in the next section.

\section{Non-equilibrium Identities \label{sec-FT}}
The stochastic equations such as the Langevin or the Fokker-Planck equations
describe how a system is dynamically 
relaxed to a stationary or an equilibrium  state. 
Furthermore we can calculate transition amplitudes of a system
to one state to another. 
By using the method reviewed in the previous section, 
we can calculate a ratio of an entropy decreasing 
probability to an entropy increasing probability.
Since the latter probabilities have always much bigger values,
the entropy is always increasing after we take a stochastic average.

In this section we review a derivation of 
the fluctuation theorem and the Jarzynski equality
from the stochastic equations.


\subsection{The Fluctuation Theorem }
The fluctuation theorem was first discovered in a numerical simulation
\cite{Evans:1993} and gives 
the ratio of probabilities of 
an entropy increasing process to that of a decreasing one.
The proof of the fluctuation theorem
is given for various systems
including classical Hamiltonian dynamics \cite{Evans:2002}, stochastic Langevin dynamics \cite{Hatano:2001} 
and quantum mechanical evolutions \cite{PhysRevA.61.062314, Kurchan:2000}.
The Jarzynski equality \cite{Jarzynski:1997}
is a relation between non-equilibrium work and equilibrium free energy difference,
and both of them are 
remarkable discoveries in the recent developments of non-equilibrium statistical physics.
In this paper, we concentrate on a system that the evolution is described by a Fokker-Planck equation such as \text{eq}.(\ref{FP}).
The fluctuation theorems
can be simply derived and the meaning of entropy production 
(or a violation of time-reversal symmetry) is clear.

We consider a stochastic system described by the Langevin equation
(\ref{Langevin})  or the Fokker-Planck equation (\ref{FP}).
In order to study a dynamical evolution, we introduce an
externally controlled parameter $\lambda^F_t$ in the 
potential $V(x; \lambda^F_t)$.
By changing the external
 parameter $\lambda^F_t$  as a function of $t$, the corresponding
 stable state changes accordingly with time.
For later convenience, we call the process of changing 
the external parameter with $\lambda^F_t$ as 
the "forward protocol". 
For example, 
we may set the minimum position of a harmonic potential 
as the externally controlled parameter;
\begin{align}
V(x;\lambda_t^F)=\frac{1}{2}k(x-\lambda_t^F)^2, 
\end{align}
if the position moves linearly  in time $t$, the parameter
is given by $\lambda_t^F=v_0 t.$
We can also take different protocols  e.g. oscillatory or pulse-like etc. 

From the path integral representation of the transition rate (\ref{Path}), 
a probability that  a sequence of configurations 
$\Gamma_\tau= \{ x(t), t\in [0,\tau] |  x(0)=x_{\text{ini}}, x(\tau)=x_{\text{fin}} \}$ is realized
during the time interval $t\in [0,\tau]$ is given by 
\begin{align}
P^F[\Gamma_\tau|x_{\text{ini}}]&\propto \exp \left[-\tfrac{1}{4\gamma T}\int_{\Gamma_\tau}dt
\left(m\ddot{x}+\gamma \dot{x} +\tfrac{\partial V(x;\lambda^F_t)}{\partial x}\right)^2\right].
\end{align}
The trajectory $\Gamma_\tau$ represents  a sequence of configurations 
in the forward protocol $\lambda^F_t$ with 
the initial configuration $x(0)=x_{\text{ini}}$.

We now define a time reversal of the forward protocol $\lambda^F_t$,
and call it a "reversed protocol" $\lambda^R_t\equiv \lambda^F_{\tau-t}$.
We consider a probability $P^R[\Gamma_\tau^\ast|x_{\text{fin}}]$ 
that the system experiences a reversed trajectory 
$\Gamma_\tau^\ast= \{ x^\ast(t)\equiv x(\tau-t), t\in [0,\tau] |  x^\ast(0)=x_{\text{fin}}, x^\ast(\tau)=x_{\text{ini}} \}$
in the time-reversed protocol $\lambda^R_t$. 
The reversed trajectory has
 the initial value $x^\ast(0)=x_{\text{fin}}=x(\tau)$, 
$\dot{x}^\ast(0)=-\dot{x}(\tau)$.
If the system has time-reversal symmetry, the probability should be the same
as the probability $P^F[\Gamma_\tau|x_{\text{ini}}]$.
 But since the stochastic equation violates the symmetry, they will be different.
The reversed propability 
$P^R[\Gamma_\tau^\ast|x_{\text{fin}}]$ is similarly given by
\begin{align}
P^R[\Gamma_\tau^\ast|x_{\text{fin}}]&\propto \exp \left[-\tfrac{1}{4\gamma T}\int_{\Gamma_\tau^\ast}dt
\left(m\ddot{x}+\gamma \dot{x} +\tfrac{\partial V(x ; \lambda^R_t)}{\partial x}\right)^2\right]\nn
&=\exp \left[-\tfrac{1}{4\gamma T}\int_{\Gamma_\tau}dt'
\left(m\ddot{x}-\gamma \dot{x} +\tfrac{\partial V(x; \lambda^F_{t'})}{\partial x}\right)^2\right].
\end{align} 
In the last line, we change a variable from 
$t$ to $t'=\tau-t$. This change causes a flip of the sign of $\dot{x}$.
The ratio of $P^F$ and $P^R$ now becomes
\begin{align}
\frac{P^F[\Gamma_\tau|x_{\text{ini}}]}{P^R[\Gamma_\tau^\ast|x_{\text{fin}}]} &=\exp \left[
-\tfrac{1}{T}\int_{\Gamma_\tau}dt\dot{x}\left(m\ddot{x} +\tfrac{\partial V(x; \lambda^F_t)}{\partial x}\right) \right] .
\label{FRratio}
\end{align}
This gives a key property to prove the fluctuation theorem.
Time-reversal symmetric terms are 
canceled  between $P^F$ and $P^R$,
and the ratio  is given by 
the entropy production $\dot{S}$ of the stochastic process. 

We further need to sum over the initial configurations,
$x_{\text{ini}}$ and $x_{\text{fin}}$ respectively
for the forward and the reversed protocols,
with appropriate statistical weights.
Here we assume that 
the external parameter is kept fixed at the initial value
of each protocol before $t=0$. 
Hence the system is in the equilibrium.
We therefore multiply $P^F$ or $P^R$ by the Boltzmann weight
$P^{\text{\text{eq}}}(x_{\text{ini}})$ or $P^{\text{\text{eq}}}(x_{\text{fin}})$. 
The ratio of the Boltzmann weights  for the initial configurations
is given by 
\begin{align}
\frac{P^{\text{eq}}(x_{\text{ini}})}{P^{\text{eq}}(x_{\text{fin}})}
&=\frac{Z(\lambda_\tau^F)}{Z(\lambda_0^F)}
\exp\left[-\frac{1}{T}\left(\frac{1}{2}m(\dot{x}_{\text{ini}}^2-\dot{x}_{\text{fin}}^2)+V(x_{\text{ini}};\lambda_0^F)-
V(x_{\text{fin}};\lambda_\tau^F)\right) \right]\nn
&=\exp\left[\frac{1}{T}\int_{\Gamma_\tau} dt \left(m\dot{x}\ddot{x}+\dot{x}\frac{\partial V(x;\lambda_t^F)}{\partial x}+
\dot{\lambda}_t^F\frac{\partial V(x;\lambda_t^F)}{\partial \lambda_t^F} \right)-\frac{\Delta F}{T}\right],
\label{BWratio}
\end{align}
where  $\Delta F$ is a difference of the free energies $ F(\lambda)= -T \log Z(\lambda)$
of equilibrium states at $\lambda=\lambda^F_0$ and
  $\lambda=\lambda^F_\tau$, 
\begin{align}
\Delta F = F(\lambda_\tau^F)-F(\lambda_0^F).
\end{align}

Combining the two ratios \text{eq}.(\ref{FRratio}) and \text{eq}.(\ref{BWratio}),
we  get the following relation,
\begin{align}
\frac{P^F[\Gamma_\tau|x_{\text{ini}}] P^{\text{eq}}(x_{\text{ini}})}{P^R[\Gamma_\tau^\ast|x_{\text{fin}}]P^{\text{eq}}(x_{\text{fin}})} &
=\exp \left( { R[\Gamma_\tau] } \right).
\label{Key}
\end{align}
Here we have defined  $R[\Gamma_\tau]$ and $W[\Gamma_\tau]$ as
\begin{align}
R[\Gamma_\tau]\equiv 
\frac{1}{T}\int_{\Gamma_\tau}dt \dot{\lambda}_t^F \frac{\partial V(x;\lambda_t^F)}{\partial \lambda_t^F} -\frac{\Delta F}{T}
 \equiv W[\Gamma_\tau] -\frac{\Delta F}{T}
\end{align}
which measures the entropy production in the trajectory $\Gamma_\tau$
and the work exerted on the system.

\begin{figure}[ht]
\begin{center}
\begin{overpic}[scale=1.0]{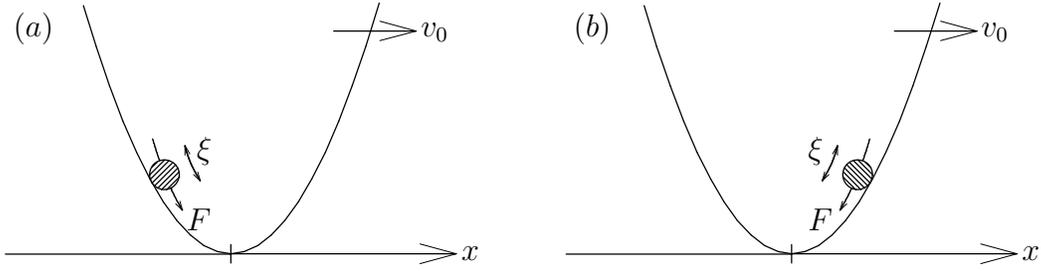}
\put(1,23){$(a)$}
\put(41,23){$v_0$}
\put(19,11){$\xi$}
\put(18,4){$F$}
\put(45,1){$x$}
\put(56,23){$(b)$}
\put(96,23){$v_0$}
\put(79,11){$\xi$}
\put(79,4){$F$}
\put(100,1){$x$}
\end{overpic}
\caption{(a) A schematic illustration of motion of a particle in a potential 
$V(x;\lambda_t^F)=\frac{1}{2}k(x-v_0 t)^2$.
This picture shows a {\it natural} configuration with $(x(t)-v_0 t) <0$.
It gives a positive value of $R[\Gamma_\tau]$.
(b) A noise $\xi$ rarely pushes a particle to the opposite side
beyond the minimum point $x(t)=v_0 t$.
Since $(x(t)-v_0 t) >0$,
it gives a negative value of $R[\Gamma_\tau]$}
\end{center}
\end{figure}

As a simple example, for the potential $V(x;\lambda_t^F)=k(x-v_0 t)^2/2$,
we have
\begin{align}
 R[\Gamma_\tau]
=-
\frac{1}{T}\int_{\Gamma_\tau}dt v_0 k(x(t)-v_0 t).
\end{align}
The term, velocity times force, gives a work exerted on the system.
If we neglected the fluctuation of the particle, 
$x(t)-v_0 t$ 
would always have a negative sign, 
and $R[\Gamma_\tau]$ would always increase.
It is consistent with a naive picture.
However in a mesoscopic system, 
fluctuations can grow larger and $x(t)-v_0 t$ can have a positive sign.
Then the particle overshoots 
the equilibrium point $\partial_x V=0$ to the positive side
and $R[\Gamma_\tau]$ becomes negative. 
Such a negative value of $R[\Gamma_\tau]$ indicates that 
the system exerts work onto outside and 
it gives a negative entropy production.

From the equation (\ref{Key}), by integrating all the paths of the configurations,
we can derive the fluctuation theorem 
in the final form as
\begin{align}
\rho^F(R_\tau) &\equiv \int {\cal D}x P^F[\Gamma_\tau|x_{\text{ini}}]P^{\text{eq}}(x_{\text{ini}}) \delta(R_\tau-R[\Gamma_\tau])\nn
&=\int {\cal D}x P^R[\Gamma_\tau^\ast|x_{\text{fin}}]P^{\text{eq}}(x_{\text{fin}})e^{R[\Gamma_\tau]} \delta(R_\tau-R[\Gamma_\tau])\nn
&=e^{R_\tau}\int {\cal D}x P^R[\Gamma_\tau^\ast|x_{\text{fin}}]P^{\text{eq}}(x_{\text{fin}}) \delta(R_\tau+R[\Gamma_\tau^\ast])\nn
&=e^{R_\tau} \rho^R(-R_\tau).
\label{FT}
\end{align}
The first line is the definition of $\rho^F(R_\tau)$, i.e. the 
probability to get the entropy production $R_\tau$
within the interval $[0,\tau]$.
We use the relation (\ref{Key}) in the second line.
In the third equality the relation $R[\Gamma_\tau^\ast]=-R[\Gamma_\tau]$ is used.
Since the quantity $R_\tau$ measures the entropy production in the interval, 
we see that  entropy decreasing probabilities are related 
to increasing ones.
They are exponentially suppressed, but exist with nonzero probabilities.

\subsection{The Jarzynski Equality}
By integrating the fluctuation theorem over the entropy production, 
we can construct an equality, 
so called the Jarzynski equality \cite{Crooks:1998}.
\begin{align}
\int_{-\infty}^\infty d R_\tau\rho^F(R_\tau) e^{-R_\tau} &= \int_{-\infty}^\infty d R_\tau\rho^R(-R_\tau)\nn
\Rightarrow \langle e^{-R_\tau} \rangle &= 1.
\label{Jarzynski}
\end{align}
We have defined the average as 
\begin{align}
\langle F(R_\tau) \rangle &= \int_{-\infty}^\infty d R_\tau\rho^F(R_\tau) F(R_\tau)
=\int {\cal D}x P^F[\Gamma_\tau|x_{\text{\text{ini}}}]P^{\text{eq}}(x_{\text{\text{ini}}}) F(R[\Gamma_\tau]) .
\end{align}
The Jarzynski equality (\ref{Jarzynski}) states that
the weighted sum of $e^{-R_\tau}$ 
over all possible non-equilibrium processes with 
an externally controlled potential gives an unity.
In terms of the work exerted on the system $W[\Gamma_\tau]$ and
the free energy difference,
we can relate an average work done in  non-equilibrium processes 
to the equilibrium free energy difference \cite{Jarzynski:1997} as
\begin{align}
\langle e^{-\frac{W}{T} } \rangle =e^{-\frac{\Delta F}{T}}.
\end{align}
From this, by using the Jensen 
inequality $\langle e^x \rangle \geq e^{\langle x \rangle}$,  we get
an inequality;
\begin{align}
\langle W\rangle -\Delta F\geq 0.
\end{align}
This indicates the second law of thermodynamics. 
The Jarzynski equality simply states that 
there must exist microscopic processes with a
large negative entropy production to satisfy the equality,
and the probability is
 characterized by the equilibrium quantity of the free energy difference. 
 
Some comments are in order. First  
the notion of entropy is usually defined  for a thermal system after taking
an average.
So it may be appropriate to use a word,
an entropy function, instead of the entropy for 
each microscopic configuration.
The second comment is that in the above {\it derivation} of the second law
we have implicitly in mind that 
the above free energy difference is the difference between the initial and 
the final free energies.
It is justified if the system is relaxed to an
equilibrium state with the external parameter at $t=\tau$ 
after a long time interval.
Since the system is  in contact with a large heat bath with temperature $T$,
the relaxed state coincides with the equilibrium state at the temperature.
If this is the case, the second law
of thermodynamics is derived from the Jarzynski  equality.
In the present proof of the fluctuation theorem, we have
used the stochastic approach and the system explicitly 
violates the time-reversal symmetry. Then such a  relaxation can 
occur. But if we starts from the original unitary quantum
mechanical evolution,
the system cannot be thermalized  in an exact sense.
In applying the fluctuation theorem 
to the information paradox of a black hole,
such considerations are inevitable.
\\

An alternative expression of the fluctuation theorem is obtained by using a generating function.
We define the generating function for $R_\tau$ as
\begin{align}
Z^F(\alpha_\tau)&= \ln\left(\int_{-\infty}^\infty d R_\tau e^{i\alpha_\tau R_\tau} \rho^F(R_\tau)\right).
\end{align}
Derivatives of $Z^F(\alpha_\tau)$ give connected correlators of 
the entropy production $R_\tau$ in a situation of the forwardly varying parameter.
One easily gets the following 
relation between $Z^F(\alpha_\tau)$ and $Z^R(\alpha_\tau)$ from the fluctuation theorem as
\begin{align}
Z^F(\alpha_\tau)&= \ln\left(\int_{-\infty}^\infty d R_\tau e^{i\alpha_\tau R_\tau}e^{R_\tau} \rho^R(-R_\tau)\right)\nn
&=\ln\left(\int_{-\infty}^\infty d x e^{ix(i-\alpha_\tau)} \rho^R(x)\right)\nn
&=Z^R(i-\alpha_\tau).
\end{align}
We have used the equation (\ref{FT}) in the first line.
In the second line, we changed a variable $R_\tau$ to $x=-R_\tau$.
If the forward  and the reversed protocols are identical i.e. $\lambda^F_t =\lambda^F_{\tau-t}$, 
we get a simpler relation $Z(\alpha_\tau)=Z(i-\alpha_\tau)$.

Finally, we give a 
 comment on our assumption of the initial distribution.
We have assumed that the initial distribution is an equilibrium one.
This condition can be easily relaxed to a steady state.
More generally, if  the initial distributions for $x_{\text{ini}}$ and $x_{\text{fin}}$ are $P^{\text{st}}(x_{\text{ini}})$  and $P^{\text{st}}(x_{\text{fin}})$
respectively, 
we can define an entropy production as 
\begin{align}
R[\Gamma_\tau]\equiv \ln \left(\frac{P^F[\Gamma_\tau |x_{\text{ini}}]P^{\text{st}}(x_{\text{ini}})}{P^R[\Gamma^\ast_\tau |x_{\text{fin}}]P^{\text{st}}(x_{\text{fin}})} \right).
\end{align}
Then we get the fluctuation theorem in the form;
$\rho^F(R_\tau)/\rho^R(-R_\tau)=e^{R_\tau}$ .
The choice of initial distributions is arbitrary,
but the problem is that we usually do not know an explicit form
of the distribution function of a steady state $P^{\text{st}}$. 
The steady state fluctuation theorems are reviewed in the 
appendix \ref{A-SSFT}.
By using it, we can derive the Green-Kubo formula and its non-linear generalizations.

\section{Langevin equation in a Black Hole Background \label{sec-BH}}
In this section we derive a stochastic equation for a scalar field
in the black hole background. 
We take $\hbar \rightarrow 0$ limit
with the Hawking temperature $\hbar \kappa /2 \pi$ fixed.
Since the energy is absorbed into the black hole, 
a dissipation term is induced at the horizon. 
The classical equation is furthermore modified 
 by the quantum effect,
 i.e. the Hawking radiation from the black hole. 
Because of these effects, the equation of motion in the black hole 
background is described by a stochastic
Langevin equation with a quantum noise and a classical dissipation terms.
We first review the basics of black holes and the 
Hawking radiation, and then derive the Langevin equation of a scalar field 
in the black hole background.
\subsection{Space-time Structure of Black Holes}
First we summarize some basic facts of the  space-time structure of black holes.
(For a review, see for example \cite{Hawking:1975e}.)
Here, we consider a
spherically symmetric neutral black hole, the Schwarzschild black hole.
It is a solution to the Einstein equation in vacuum with a 
zero cosmological constant
and the metric is given by 
\begin{align}
& ds^2=-f(r)dt^2+\frac{dr^2}{f(r)}+r^2d\Omega^2, \nn
& f(r)=1-\frac{2GM}{r}, \ \ \ d\Omega^2=d\theta^2+\sin^2\theta d\phi^2.
\label{Schwarz}
\end{align}
$M$ is the mass of the black hole and  the only parameter of the solution.
The solution is asymptotically flat;
it approaches the flat  metric at the space-like infinity $r\to \infty$.
It has  time-translation symmetry and the 
associated time-like Killing vector is given by $\xi=\partial_t$.
A Killing horizon is defined as
a null hypersurface on which there is a null Killing vector.
In the present case, it is given by the condition 
$g(\xi, \xi)=-f(r)=0   \leftrightarrow r=r_H=2GM$.
The surface gravity $\kappa$ is defined on the Killing horizon via the relation
\begin{align}
\nabla_{\xi}\xi=\kappa \xi.
\end{align}
A direct calculation shows that $\kappa=f^\prime(r)/2|_{r=r_H}=1/4GM$ for the Schwarzschild black hole. 

There are several different definitions of horizons.
An apparent horizon is a more general concept and defined locally
as the most outer trapped null surface.
It does not need a time-like Killing vector as the Killing horizon,
but it is defined in an observer-dependent way.
An event horizon is defined in a global way as a boundary
of the past light cone of the future infinity.
Mathematically a black hole is defined as a set that is not contained
in the past light cone of the future infinity.
For the Schwarzschild black hole, all the definitions of the horizon
coincide, but they are different for dynamical black holes.
In  applying non-equilibrium  statistical physics 
to the dynamics of black holes, we need to pay special attentions to 
the differences.
In the present paper, however, 
since we consider an eternal black hole 
as a background space-time,
their differences are not essential.

The coordinates used in \text{eq}.(\ref{Schwarz}) is called the
Schwarzschild coordinates. 
The singularity of the metric at the horizon $r=r_H$ is not physical,
and can be removed by using other coordinates, 
such as the
 Kruskal (-Szekeres) coordinates $(U, V)$
\begin{align}
U&=-\frac{1}{\kappa}e^{-\kappa(t-r_\ast)}, \ \ 
V=\frac{1}{\kappa}e^{\kappa(t+r_\ast)} 
\label{UV} \\
r_\ast&\equiv \int\frac{dr}{f(r)}=r+r_H\log(\frac{r}{r_H}-1).
\end{align}
Here $r_*$ is the tortoise coordinate and takes $-\infty <r_* < \infty$ between the 
horizon and the spacial infinity.
In terms of  the Kruskal  coordinates, 
the metric of the Schwarzschild black hole becomes regular at the horizon; 
\begin{align}
ds^2&=-\frac{r_H}{r}e^{-\frac{r}{r_H}}dUdV+r^2d\Omega^2.
\end{align}
At the price of removing the coordinate singularity, 
the asymptotically flatness is unclear in these coordinates.
We will impose regularity conditions on physical quantities
at the horizon in the Kruskal coordinates.

\begin{figure}[ht]
\begin{center}
\begin{overpic}[scale=0.6]{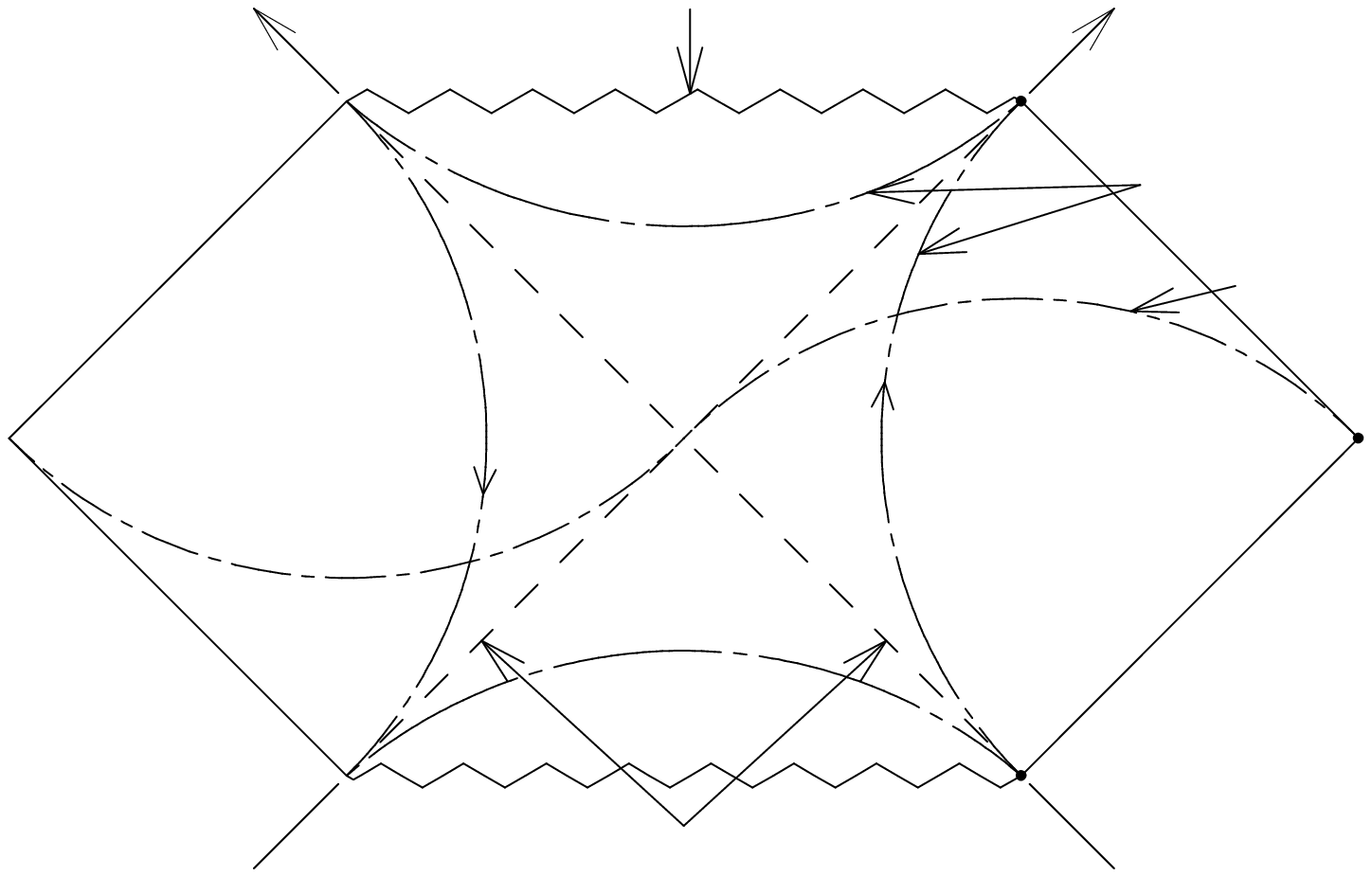}
\put(45,64){$r=0$}
\put(15,64){$U$}
\put(81,64){$V$}
\put(73,58){$i^+$}
\put(72,2){$i^-$}
\put(100,30){$i^-$}
\put(44,1){$r=r_H$}
\put(84,50){$r=$const.}
\put(91,42){$t=$const.}
\put(74,30){I}
\put(23,30){II}
\put(47,43){III}
\put(48,18){IV}
\end{overpic}
\caption{
A point in the diagram represents a two dimensional sphere with  radius $r$
at time $t$.
$r$-constant and $t$-constant surfaces are depicted.
Arrows on the $r$-constant surfaces 
indicate the flow of the time-like Killing vector.
They have  opposite directions in the region I and I\hspace{-0.1em}I.
The singularity at $r=0$ is drawn by zigzag lines in the diagram.
Event horizons are located at $r=r_H$ and separate the space-time into four distinct regions. 
$i^+$, $i^-$ and $i_0$ are the future, past and spatial infinities. 
}
\label{penrose}
\end{center}
\end{figure}

Figure \ref{penrose} is the Penrose diagram of the Schwarzschild
black hole, which captures the causal structure of the space-time.
The vertical and horizontal axises correspond to the Kruskal time $T=(V+U)/2$, 
and the  Kurskal radius $R=(V-U)/2$.
In contrast to the Schwarzschild coordinates, 
the Kruskal coordinates are regular beyond the horizon ($r=r_H$),
and can be extended to the maximally extended Schwarzschild space-time
($-\infty < U, V <\infty$).
The original Schwarzschild coordinates ($-\infty< t <\infty , r_H<r <\infty$),
on the contrary,  can cover only the  region I in fig.\ref{penrose}. 
We define $(t,r_\ast)$ coordinates  in other regions.
For example, in the region 
I\hspace{-0.1em}I,
we can define them by
the relations 
$U=e^{-\kappa(t-r_\ast)}/\kappa, V=-e^{\kappa(t+r_\ast)}/\kappa$.
In the Kruskal coordinates, the space-time is separated by the future
and past event horizons ($U=0$ and $V=0$ respectively) into four regions.
There are four possible combinations of signature of $U$ and $V$
as shown in the table \ref{4regions}.
\begin{table}[t]
\begin{center}
\begin{tabular}{c|c|c}
\hline
I &\ $U<0 , V>0$\ &\ $r>r_H$ \\ \hline
I\hspace{-0.1em}I&\ $U>0 , V<0$\ &\ $r>r_H$ \\ \hline
I\hspace{-0.1em}I\hspace{-0.1em}I&\ $U>0 , V>0$\ &\ $r<r_H$ \\ \hline
I\hspace{-0.1em}V&\ $U<0 , V<0$\ &\ $r<r_H$ \\
\hline
\end{tabular}
\caption{Four regions of maximally extended Schwarzschild space-time} 
\end{center}
\label{4regions}
\end{table}

Finally we note that the time-like Killing vector $\xi=\partial_t$
is written as $\xi=\kappa(V\partial_V-U\partial_U)=\kappa R\partial_T$ in the Kruskal coordinates
and, therefore, the directions of time are opposite in the 
region I and  I\hspace{-0.1em}I.
We have drawn the directions of $\xi$ in fig.\ref{penrose}.
\subsection{Field Theory in the Black Hole Background and the Hawking Radiation}
We briefly summarize the quantum field theories in the black hole
background. For a comprehensive review, see e.g. \cite{Birrell:1984}.
The action of 
a massive scalar field in the maximally extended Schwarzschild space-time
is given by a sum of the fields in the right wedge (region I)
and in the left wedge (region I\hspace{-0.1em}I\hspace{-0.1em}I).
In each region, the action is given by 
\begin{align}
S=
\int d^4x \sqrt{-g}\frac{1}{2}
\left(g^{\mu\nu}\partial_\mu\phi \partial_\nu \phi +m^2\phi^2\right)
=
\sum_{l,m} \int dt dr_\ast \phi_{ (l,m)}\left[
\partial_t^2-\partial_{r_\ast}^2+V_l (r)  \right]\phi_{ (l,m)}.
\end{align}
where we have decomposed the field into partial waves 
\begin{align}
 & \phi(t,r,\Omega)=\sum_{l,m}\frac{\phi_{(l,m)}(t,r)}{r}Y_{l,m}(\Omega), 
\end{align}
and defined the effective potential for each partial wave with an 
angular momentum $l$,
\begin{align}
& V_l (r) =f(r)\left(\frac{l(l+1)}{r^2}+\frac{\partial_r f(r)}{r}+m^2 \right).
\end{align}
The equation of motion of the scalar field is given by
\begin{align}
\left[\partial_t^2-\partial_{r_\ast}^2+V_l (r)  \right]\phi_{R, L (l,m)}=0.
\label{EOM}
\end{align}
Both in the  asymptotically flat region ($r\to\infty$) 
and in the near horizon region ($r\to r_H$),
the potential $V_l$  vanishes and 
the equation of motion is reduced to the free field equation.
Thus,  in the near horizon region,  the classical solutions 
are approximately given by
\begin{align}
u_k^R(t, r)&= 
\left\{
\begin{array}{cc}
\frac{1}{\sqrt{4\pi\omega_k}} e^{-i\omega_k t +ik r_\ast} & (\text{in R})\\
0 &  (\text{in L})
\end{array}
\right.
\\
u_k^L(t, r)&=
\left\{
\begin{array}{cc}
0 &  (\text{in R})\\
\frac{1}{\sqrt{4\pi\omega_k}} e^{i\omega_k t +ik r_\ast} & (\text{in L}) .
\end{array}
\right. 
\label{wf-Min}
\end{align}
and their complex conjugates.
Here $\omega_k=+|k| >0$. 
The sign difference in front of $i\omega_k t$ in $R, L$ follows
the convention of \cite{Birrell:1984}.
With this convention, these fields  are positive frequency modes 
with respect to the time-like Killing vector, $\partial_t$ in R
and $-\partial_t$ in L,
satisfying ${\cal L}_{\pm \partial_t} u_k=-i\omega_k u_k$.
The complex conjugates $(u_k^{R,L})^\ast$ are 
the negative frequency modes (in the above sense)
satisfying ${\cal L}_{\pm \partial_t} u_k^\ast=+i\omega_k u_k^\ast$.
They are orthonormal with respect to the following 
Klein-Gordon inner product,
\begin{align}
(f,g)&\equiv i\int_{\Sigma_t} d^3x \sqrt{h_{\Sigma_t}}\left(f^\ast \partial_t g-\partial_t f^\ast g \right)\nn
&=i\sum_{l,m}\int dr_\ast \left(f_{(l,m)}^\ast\partial_t g_{(l,m)}-\partial_t f_{(l,m)}^\ast g_{(l,m)} \right).
\end{align}
The integration is performed  on a constant time slice $\Sigma_t$, 
but it can be generalized to any space-like surface $\Sigma$
and the choice of the integration surface does not change the value of the 
inner product. 
The field $\phi_{(l,m)}$ can be expanded 
in terms of the classical solutions in the Schwarzschild coordinates
in the near horizon region as follows;
\begin{align}
\phi_{(l,m)}&= \int\frac{d k}{\sqrt{4\pi \omega_k}}\left[a^R_{k (l,m)} u^R_k+(a^R_{k (l,m)})^{\dagger} (u^R_k)^\ast
+a^L_{k (l,m)} u^L_k+(a^L_{k (l,m)})^{\dagger} (u^L_k)^\ast
 \right].
\end{align}
We will omit the suffixes $(l,m)$ of creation and annihilation operators for simplicity in the followings.

In the Kruskal coordinates near the horizon, 
the equation of motion becomes $\partial_U\partial_V\phi_{(l,m)}=(\partial_T^2-\partial_R^2)\phi_{(l,m)}=0$.
So we may define another basis of functions
\begin{align}
u^K_p(T, R)&= 
\frac{1}{\sqrt{4\pi E_p}} e^{-iE_p T +ip R} ,
\end{align}
where $E_p=+|p|>0$.  They are positive frequency modes
with respect to  the Kruskal time. 
In terms of them, the field can be expanded as 
\begin{align}
\phi_{(l,m)}&= \int\frac{d p}{\sqrt{4\pi E_p}}\left[b_p u^K_p+(b_p)^{\dagger} (u^K_p)^\ast
 \right].
\end{align}
In contrast to the wave functions (\ref{wf-Min}), they are
defined globally in the whole space-time.

In order to relate two different definitions of the creation and annihilation
operators in the Kruskal and Schwarzschild coordinates
and to express the Kruskal vacuum $b_k|0\rangle_K=0$
as a  Fock state constructed on the Schwarzschild vacuum
$a^{R,L}_k|0\rangle_{R, L}=0$, we look at the analyticity properties
of the functions \cite{Unruh:1976db}.
The positive frequency wave function $u^K_p$ in the Kruskal coordinates
with $p>0$ (or $p<0$) 
is an analytic function in the lower half $U$ (or $V$) plane 
since $u^K_p \sim e^{-iE_p U}$ (or $u^K_p \sim e^{-iE_p V}$).
On the other hand, 
since $e^{ik r_\ast}=(r/r_H-1)^{ik}e^{ik r}$, there is a phase jump
when it crosses the horizon.
So we need to combine the positive and negative 
frequency wave functions in the Schwarzschild coordinates
to construct a wave function with the same analyticity property as $u^K_p$.
They were obtained by Unruh \cite{Unruh:1976db} as
\begin{align}
\left\{
\begin{array}{c}
u_k^{(1)}=\frac{1}{\sqrt{2\sinh \frac{\pi \omega_k}{\kappa \hbar}}}\left(
e^{\frac{\pi\omega_k}{2\kappa \hbar}} u_k^R + e^{-\frac{\pi \omega_k}{2\kappa \hbar}}(u^L_{-k})^\ast \right)\\
u_k^{(2)}=\frac{1}{\sqrt{2\sinh \frac{\pi \omega_k}{\kappa \hbar}}}\left(
e^{-\frac{\pi\omega_k}{2\kappa \hbar}} (u_{-k}^R)^\ast + e^{\frac{\pi \omega_k}{2\kappa \hbar}}u^L_{k} \right).
\end{array}
\right.
\label{Unruh}
\end{align}
These combinations are analytic in the lower half plane of $U$ or $V$.
In the following we set $\hbar=1$ for notational simplicity.
Such analyticity property can be easily checked.
For example, $u_k^{(1)}$ with a positive $k$ can be rewritten as
an analytic function of $U$
\begin{align}
u^{(1)}_k&\propto u_k^R+e^{-\frac{\pi\omega_k}{\kappa}}(u^L_{-k})^\ast \nn
& \propto (-\kappa U)^{\frac{i\omega_k}{\kappa}} ,
\end{align} 
if it is analytically continued
from the region I of the right wedge ($U<0$) to the region I\hspace{-0.1em}I 
of the left wedge ($U>0$)
through the lower half of the $U$ plane by
the transformation $U \rightarrow U e^{i\pi}$.
Hence the combination is analytic in the lower half plane of $U$.
For $k<0$, 
$ u^{(1)}_k \propto (\kappa V)^{-\frac{i\omega_k}{\kappa}} $ and
it is analytic in the lower half plane of $V$  as $e^{-iE_p V}$.
In the classical limit where $\hbar \rightarrow 0$, 
$u_k^{(1)}$ 
 becomes a positive frequency mode in the Schwarzschild
coordinates $e^{-i\omega_k (t \mp r_\ast)}$ and  localized in the region I.
Similarly, $u_k^{(2)}$ with a positive momentum $k>0$
is written as an analytic function of the lower half plane of $V$,
$(\kappa V)^{i\omega_k/\kappa}$ while, for a negative $k$, 
it is analytic in the lower half plane of $U$ and written as
 $(-\kappa U)^{-i\omega_k/\kappa}$.
It behaves as a negative frequency mode in the Schwarzshchild coordinates
but localized  mostly in the left wedge in the classical limit. 
They penetrate into the right wedge by quantum effects.
In this sense, $u_k^{(1)}$ is {\it classical} while $u_k^{(2)}$
is  {\it quantum} in the right wedge.

The scalar field can be expanded in terms of 
 these modes as
\begin{align}
\phi_{(l,m)}&= \int\frac{d k}{\sqrt{4\pi \omega_k}}\left[c^{(1)}_k u^{(1)}_k+(c^{(1)}_k)^{\dagger} (u^{(1)}_k)^\ast
+c^{(2)}_k u^{(2)}_k+(c^{(2)}_k)^{\dagger} (u^{(2)}_k)^\ast \right].
\end{align}
The Kruskal vacuum ($b_p|0\rangle_K=0$) is equivalently
given by the conditions,
$c^{(1)}_k|0\rangle_K=c^{(2)}_k|0\rangle_K=0$.
The annihilation operators in the Schwarzschild coordinates
$a^R_k$ and $a^L_k$ can be expressed as a linear combination of $c^{(1)}_k$ and $c^{(2)}_k$
as
\begin{align}
\left\{
\begin{array}{cc}
a^R_k
=\frac{1}{\sqrt{2\sinh \frac{\pi \omega_k}{\kappa}}}\left(
e^{\frac{\pi\omega_k}{2\kappa}} c^{(1)}_k + e^{-\frac{\pi \omega_k}{2\kappa}}(c^{(2)}_{-k})^\dagger \right) 
&=\sqrt{1+n(\omega_k)}c^{(1)}_k+\sqrt{n(\omega_k)}(c^{(2)}_{-k})^\dagger
\\
a^L_k
=\frac{1}{\sqrt{2\sinh \frac{\pi \omega_k}{\kappa}}}\left(
e^{\frac{\pi\omega_k}{2\kappa}} c^{(2)}_k + e^{-\frac{\pi \omega_k}{2\kappa}}(c^{(1)}_{-k})^\dagger \right) 
&=\sqrt{1+n(\omega_k)}c^{(2)}_k+\sqrt{n(\omega_k)}(c^{(1)}_{-k})^\dagger.
\end{array}
\right.
\end{align}
where $n(\omega_k)=1/(e^{2\pi \omega_k/\kappa} -1)$.
Hence the Kruskal  and the Schwarzschild operators are related
by the Bogoliubov transformation,
\begin{align}
\left(
\begin{array}{c}
a_k^R\\
(a_{-k}^L)^\dagger
\end{array}
\right)
&=
\left(
\begin{array}{cc}
\sqrt{1+n(\omega_k)} & \sqrt{n(\omega_k)}\\
\sqrt{n(\omega_k)} & \sqrt{1+n(\omega_k)}
\end{array}
\right)
\left(
\begin{array}{c}
c_k^{(1)}\\
(c_{-k}^{(2)})^\dagger
\end{array}
\right)
\equiv U_k
\left(
\begin{array}{c}
c_k^{(1)}\\
(c_{-k}^{(2)})^\dagger
\end{array}
\right)
.
\label{bogoliubov}
\end{align}
The transformation can also be represented as
\begin{align}
c^{(1)}_k&= e^{-iG}a^R_k e^{iG},\ \ 
c^{(2)}_{-k}= e^{-iG}a^L_{-k} e^{iG},\nn
G&=i\int \frac{dk}{(2\pi)2\omega_k}\theta_k\left(
(a_k^R)^\dagger(a_{-k}^L)^\dagger-a_k^R a_{-k}^L
\right),\nn
\sinh^2\theta_k&\equiv n(\omega_k).
\label{relation_vac}
\end{align}
From this transformation law, 
we can read off the relation between Kruskal vacuum and Schwarzschild vacuum as
\begin{align}
|0\rangle_K&= e^{-iG}|0\rangle_R |0\rangle_L\\
&=\prod_k \frac{1}{\cosh \theta_k}\sum_{n=0}^\infty e^{-\frac{\beta \omega_k}{2} n_k}|n_k^R\rangle |n^L_{-k}\rangle.
\end{align}
Note that the Fock space in the left wedge $|n^L_k \rangle$ is constructed
on a Minkowski vacuum with the backward time direction $(-t)$. 

The expectation value of the Schwarzschild number operators 
$(a^{R}_k)^\dagger a^{R}_k$ in the Kruskal vacuum $|0\rangle_K$
is given by 
\begin{align}
_K\langle0 | (a^R_k)^\dagger a^R_k|0\rangle_K
&=\frac{1}{2\sinh \frac{\pi \omega_k}{\kappa}} e^{-\frac{\pi\omega_k}{\kappa}}\  _K\langle0 | c^{(2)}_{-k} (c^{(2)}_{-k})^\dagger|0\rangle_K\nn
&=\frac{1}{e^{\frac{2\pi\omega_k}{\kappa}}-1} =n(\omega_k).
\end{align}
This is the thermal distribution  of the Hawking radiation 
 \cite{Hawking:1974sw}, 
and  characterized by the temperature $T_H=\kappa \hbar/2\pi$.
Note that the thermal spectrum in the right wedge is created
by the effect of the field $u_k^{(2)}$, which is classically
localized in the left wedge but penetrates into the right
quantum mechanically. 

For a generic operator $\hat{\cal O}_R=\hat{\cal O}_R (a^R, (a^R)^\dagger)$
 which is  made of only $a^R$ and $(a^R)^\dagger$,
its expectation value $_K \langle 0| \hat{\cal O}_R |0\rangle_K$
can be interpreted as a thermal average.
Such a thermal behavior can be generalized to  
 products of operators,
such as $_K \langle 0| \hat{\cal O}_L\hat{\cal O}_R |0\rangle_K$,
 made of both the right and left creation (annihilation) operators.
Its expectation value can be interpreted as a Schwinger-Keldysh
correlator.

First let us consider $_K \langle 0| \hat{\cal O}_R |0\rangle_K$.
Since the Kruskal vacuum is represented as (\ref{relation_vac}), one has
\begin{align}
_K \langle 0| \hat{\cal O}_R |0\rangle_K
&=\prod_{k } \frac{1}{\cosh^2\theta_k} \sum_{n=0}^\infty \langle n_k^R | e^{-\beta \omega_k n_k}\hat{\cal O}_R |n_k^R\rangle\nn
&=\text{Tr}_R \left[\frac{e^{-\beta H_R}}{Z_R} \hat{\cal O}_R \right].
\end{align}
Here, the Hamiltonian and the partition function are defined by
\begin{align}
H_R&=\int\frac{dk}{2\pi}\omega_k (a_k^R)^\dagger a_k^R,\nn
Z_R&=\text{Tr}\left[ e^{-\beta H_R}\right]
=\prod_k\sum_{n=0}^\infty e^{-\beta \omega_k n^R_k}
=\prod_k \cosh^2\theta_k.
\end{align}
Hence $_K \langle 0| \hat{\cal O}_R |0\rangle_K$ can be interpreted as 
a thermal average of 
the operator $\hat{\cal O}_R $ at the Hawking temperature $T_H$.

For a product of left and right operators, the expectation value in 
the Kruskal vacuum is given by
\begin{align}
_K \langle 0| \hat{\cal O}_L\hat{\cal O}_R |0\rangle_K
&=\prod_{k , k'}\frac{1}{\cosh\theta_k \cosh\theta_{k'}} \sum_{m,n=0}^\infty\ 
e^{-\frac{\beta}{2}(\omega_km_k+\omega_{k'}n_{k'})}\langle m_k^R | \langle m_{-k}^L| 
\hat{\cal O}_L\hat{\cal O}_R |n^R_{k'}\rangle|n^L_{-k'}\rangle\nn
&=\prod_{k , k'}\frac{1}{\cosh\theta_k \cosh\theta_{k'}} \sum_{m,n=0}^\infty\ e^{-\frac{\beta}{2}(\omega_km_k+\omega_{k'}n_{k'})}
\langle m_k^R | \hat{\cal O}_R|n^R_{k'}\rangle \langle m_{-k}^L| \hat{\cal O}_L |n^L_{-k'}\rangle.
\end{align}
In order to express it as an expectation value in the right 
wedge Fock space,
we first rewrite
the expectation value $\langle n_{-k}^L| \hat{\cal O}_L |m^L_{-k'}\rangle$
in terms of the operator in the right wedge 
as follows,
\begin{align}
\langle m_{-k}^L| \hat{\cal O}_L |n^L_{-k'}\rangle&=
\langle n_{k'}^R| \hat{\cal O}_L^{\vee} |m^R_{k}\rangle,
\end{align}
Here we have defined the operator  
$\hat{\cal O}_L^{\vee}(a_R,a_R^\dagger)$ by the following
substitution,
\begin{align}
\hat{\cal O}_L(a^L,(a^L)^\dagger)=\sum_n c_{m,n}  (a_k^L)^m (a_k^{L \dagger})^n
\rightarrow 
\hat{\cal O}_L^{\vee}(a^R , a^{R \dagger})\equiv 
\sum_n c_{m,n} (a_{-k}^R)^n  (a_{-k}^{R \dagger} )^m.
\label{substitution}
\end{align}
Note that the coefficients $c_{m,n}$ are not converted 
to its complex conjugate.
In particular, the field itself $\phi_L(t)$ is converted as
\begin{align}
 \phi_L(t) &= \int \frac {dk}{4\pi\omega_k}
 [a^L_k e^{i\omega_k t +ikr_\ast} +(a_k^L)^\dagger e^{-i\omega_k t -ikr_\ast}]\\
 \to \phi_L^\vee(t) &= 
 \int \frac {dk}{4\pi\omega_k}
 [(a^R_{-k})^\dagger e^{i\omega_k t +ikr_\ast} +a_{-k}^R  e^{-i\omega_k t -ikr_\ast}]. 
\end{align}
This has the same functional form as $\phi^R(t)$.
We later interpret this field $\phi_L^\vee(t)$ as 
the (lower) Schwinger-Keldysh field and write as $\tilde\phi_R$
to distinguish the original right-wedge field $\phi_R.$
By this substitution, the expectation value can be
written as
\begin{align}
_K \langle 0| \hat{\cal O}_R  \hat{\cal O}_L|0\rangle_K
&=\frac{1}{Z} 
\text{Tr} 
\left( 
e^{-\frac{\beta}{2}H_R} \ \hat{\cal O}_R \ e^{-\frac{\beta}{2}H_R}
\ \hat{\cal O}_L^\vee
\right) \\ \nonumber
& \equiv \langle \hat{\cal O}_R  \hat{\cal O}_L^\vee 
\rangle_{\frac{\beta}{2},\frac{\beta}{2}}
\end{align}
In order to distinguish it from
the ordinary finite temperature
Green function, we have introduced the notation 
$\langle \cdots 
\rangle_{\frac{\beta}{2},\frac{\beta}{2}}$ as above.

If $\hat{\cal O}_L$ is made of
a product of operators  $\hat{\cal O}_L= \hat{A}_L \hat{B}_L $, it is 
converted as
\begin{align}
\hat{\cal O}_L= \hat{A}_L \ \hat{B}_L \to 
\hat{\cal O}_L^\vee=
\hat{B}_L^\vee \ \hat{A}_L^\vee.
\end{align}
Especially a  care should be taken for the time evolution operator
$U_L(t,t_0)\equiv T\exp\Big[-i\int_{t_0}^t dt \hat{H}_L(t) \Big]$.
Following the above substitution rule,
it
is converted to 
\begin{align}
U_L(t,t_0)&\equiv T\exp\left[-i\int_{t_0}^t dt \hat{H}_L(t) \right]
\to U^\vee_L(t,t_0)= \tilde{T}\exp\left[-i\int_{t_0}^t dt 
\hat{H}_L^\vee(t) \right]
\end{align}
where $\tilde{T}$ is an anti-time ordering.
For a hermitian  Hamiltonian,  
 $\hat{H}_L^\vee = \hat{H}_R$ is satisfied
and
\begin{align}
U^\vee_L(t,t_0)= \tilde{T}\exp\left[-i\int_{t_0}^t dt 
\hat{H}_R(t) \right]
\end{align}
Hence a  Heisenberg operator $ \hat{A}_L(x)$ is mapped to 
\begin{align}
 \hat{A}_L(x)= U^\dagger_L(t_x,t_0) \hat{A}_L(t_0) U_L(t_x,t_0)
\to \hat{A}_L^\vee (x)=
U_L^\vee(t_x,t_0)  \hat{A}_L^\vee(t_0) U_L^{\dagger \vee}(t_x,t_0).
\end{align}
The converted Heisenberg operator is evolved backward in time
with the Hamiltonian $(-H_R)$. 

\begin{figure}[ht]
\begin{center}
\begin{overpic}[scale=0.6]{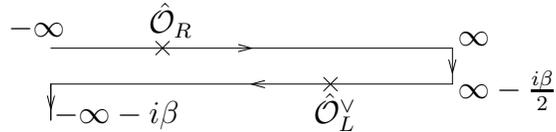}
\put(0,21){\hspace{-1.2em}$-\infty$}
\put(25,22){$\hat{\cal O}_R$}
\put(2,0){$-\infty-i\beta$}
\put(100,19){$\infty$}
\put(100,6){$\infty-\frac{i\beta}{2}$}
\put(65,0){$\hat{\cal O}^\vee_L$}
\end{overpic}
\end{center}
\caption{
$\phi_R$ lives on the upper line while 
$\tilde\phi_R$ lives on the
lower line. The time evolution is backward on the lower line.
}
\label{path_aa}
\end{figure}

From these considerations, 
an expectation value of a general operator including both of
 left and right operators
can be represented  as a path integral form of the
right-handed fields;
\begin{align}
 _K \langle 0|   \hat{\cal O}_R \hat{\cal O}_L|0\rangle_K
&= \langle    \hat{\cal O}_R \hat{\cal O}_L^\vee 
\rangle_{\frac{\beta}{2},\frac{\beta}{2}}
\nn
&=
\int {\cal D}\phi_R {\cal D} \tilde\phi_R \ {\cal O}_R[\phi_R] \ 
{\cal O}^\vee_L[\tilde\phi_R]  
\exp\left[iS[\phi_R]-iS[\tilde\phi_R]
\right].
\label{LR-SK}
\end{align}
Here  $\phi_R$ represents the original right-wedge field
while  a new field $\tilde\phi_R(t)$ is introduced 
to represent the transformed operator ${\cal O}_L^\vee$. 
The minus sign in front of the action
$S[\tilde\phi_R]$ comes from the backward time-evolution of 
${\cal O}_L^\vee$.
If we combine $\phi_R(t)$ and $\tilde\phi_R(t)$ 
together as a single $\phi_R(t)$
field along a doubled path depicted below, this expression is equivalent to the 
closed time path formalism of the real-time finite temperature field theory. 
The insertions of $\exp(-\beta H_R/2)$ can be represented
as an evolution of time into the imaginary direction with $-\beta/2$
at both ends.
Hence the path 
is given on the complex time plane as Fig. \ref{path_aa}.
The field on the lower line corresponds to the field 
in the left-wedge as 
$\phi_R(t-i\beta/2)=\phi_L(t)$.

An alternative interpretation is an analogy
 with the thermo field dynamics \cite{Takahasi:1974zn},
another method to deal with the real-time finite temperature field theory.
In this analogy,
the operators in the left wedge can be
 regarded as the "tilde-fields" of thermo field dynamics \cite {Israel:1976ur}.


\subsection{Effective Equation of Motion in the Vicinity of the Horizon}
Now we derive an effective equation of motion for the scalar field
in the black hole background. 
Classically a dissipation term is induced since the energy
is absorbed into the black hole horizon.
In quantizing the system, a noise term will also be induced 
because of the Hawking radiation, and the system is effectively
described by a Langevin equation.

The effect of the absorption can be 
described by imposing the ingoing boundary condition
 at the horizon $r=r_H$.
Since, in the near horizon region, the system can be described by a set of 
2-dim free fields satisfying
 $(\partial_t^2-\partial_{r_\ast}^2)\phi_{(l,m)}=0$, 
the ingoing boundary condition can be represented as 
\begin{align}
(\partial_t -\partial_{r_\ast})\phi_{(l,m)}(t,r=r_H)=0 .
\end{align}
The condition implies that there are no outgoing modes at the horizon,
and violates the time reversal symmetry.

Since the scalar field is coupled to the gravitational field, if it is quantized,
the chiral condition at the horizon seems to violate the general covariance
by the quantum gravitational anomaly.
The violation is compensated by the flux of the Hawking radiation
\cite{Robinson:2005pd, Iso:2006wa, Iso:2006ut}.
In the following we will see that the quantization of the scalar field
near the horizon naturally leads to the chiral condition (absorption)
with the flux of Hawking radiation (noise term) at the horizon.

The method we will use is  similar to the retarded-advanced (or Schwinger-Keldysh)
formalism. The derivation of a Langevin equation
 is given by integrating fluctuating fields.
(For a review, see, e.g. \cite{Calzetta:2008}.)  

\subsubsection{Integrating Out the  Environments }
In obtaining the Langevin equation at the horizon, 
we need to integrate  out certain kinds of 
environmental variables interacting with the {\it system} 
variable at the horizon. 
In order to do this, we first consider a stretched horizon 
at $r=r_H+\epsilon$ and treat the variables 
 between the horizon ($r=r_H$) and the 
stretched horizon ($r=r_H+\epsilon$) as the environmental variables.
Because of the quantum mixing of the wave functions in the left and right wedges
(\ref{Unruh}), the integration of these variables 
corresponds to an integration 
of the fields in the left wedge, which are classically hidden. 
In this way, we  derive a Langevin equation at the stretched horizon.
This equation  is shown to be 
independent of the small parameter $\epsilon$ characterizing the
position of the stretched horizon
and we can take $\epsilon \rightarrow 0$ limit at last.

Since the Langevin equation we are going to derive is the equation 
of motion at the boundary of a rigion $[r_H, r_H+\epsilon]$, 
it is convenient to
discretize the equation of motion near the horizon. 
In the tortoise coordinate $r_*$
in which the equation of motion becomes free, 
the region is mapped to $[-\infty, r_\ast(r_H+\epsilon)]$.
We divide the region 
 into infinite segments
as  $(r_\ast)_n=r_\ast(r_H+\epsilon) + n d $ (for $n=0, -1, -2, \cdots -\infty$)
and set oscillators  $x_n$ on these lattice points. 
Here $d$ is a lattice spacing in the tortoise coordinate.
Discretized equations of motion for the scalar field are given by
\begin{align}
\left\{
\begin{array}{l}
\ddot{x}_0 =-k(x_0-x_{1})+k(x_{-1}-x_0)- V_l((r_\ast)_0)x_0 \\
\vdots \\
\ddot{x}_{-n} =-k(x_{-n}-x_{-n-1})+k(x_{-n+1}-x_{-n})- V_l((r_\ast)_{-n})x_{-n}.
\label{disc_inner}
\end{array}
\right.
\end{align}
The continuum limit is given by taking $d \rightarrow 0$ limit with $kd^2=1$ and $\phi_{(l,m)}(t, (r_\ast)_n)=x_n(t)/\sqrt{d}$.

\begin{figure}[ht]
\begin{center}
\begin{overpic}[scale=0.6]{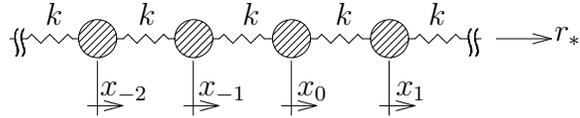}
\put(7,17){$k$}
\put(24,17){$k$}
\put(43,17){$k$}
\put(60,17){$k$}
\put(77,17){$k$}
\put(17,4){$x_{-2}$}
\put(35,4){$x_{-1}$}
\put(53,4){$x_{0}$}
\put(71,4){$x_{1}$}
\put(100,14){$r_\ast$}
\end{overpic}
\caption{Discretized model of a scalar field in the near horizon region. 
The variable $x_0$ represents a variable at the stretched horizon.}
\end{center}
\end{figure} 

Introducing the forward and the backward differentials in the
tortoise coordinate $r_\ast$,
\begin{align}
\Delta^+ x_n \equiv \frac{x_{n+1}-x_n}{d}\ , \ \Delta^- x_n\equiv \frac{x_{n}-x_{n-1}}{d},
\end{align}
and using the relation,
\begin{align}
-(x_n-x_{n+1})+(x_{n-1}-x_n)=d\left(\Delta^+ -\Delta^-\right)x_n=d^2\Delta^+\Delta^- x_n,
\end{align}
we can write the discretized equation (\ref{disc_inner}) for $n < 0$ as
\begin{align}
\ddot{\phi}_{l,m}((r_\ast)_n)&=kd^2\Delta^+\Delta^-\phi_{l,m}((r_\ast)_n)-V_l((r_\ast)_n)\phi_{l,m}((r_\ast)_n)\nn
\xrightarrow{d \rightarrow 0} \ \ddot{\phi}_{l,m}(r_\ast)&=\partial_{r_\ast}^2\phi_{l,m}(r_\ast)-V_l(r_\ast)\phi_{l,m}(r_\ast).
\end{align}
$V_l((r_\ast)_n)$ stands for the gravitational potential at $(r_\ast)_n$.
It is proportional to $ f(r)=1-2M/r$ and
vanishes in the near horizon region. Hence we neglect the 
potential term  later  in this section.

The normalization of the fields 
$\phi_{(l,m)}((r_\ast)_n)\equiv x_n/\sqrt{d}$ 
is determined from the action. 
With this normalization, the discretized action becomes the continuum one;
\begin{align}
S&=-\int dt \left[\frac{1}{2} \sum_{n=-\infty}^0 
(\dot{x}_n)^2
-U(x_n)\right] \nn
&=-\int dt \int_{-\infty}^{r_\ast(r_H+\epsilon)} dr_\ast \frac{1}{2} \left[ 
(\dot{\phi}_{(l,m)}(r_\ast))^2-
(\partial_{r_\ast} \phi_{(l,m)} (r_\ast))^2-V_l(r_\ast)(\phi_{(l,m)} (r_\ast))^2 \right].
\end{align}
Here the discretized potential $U$ is defined by
\begin{align}
U(x_n)&\equiv \frac{1}{2} \sum_{n=-\infty}^0 \left[
k (x_{n+1}-x_n)^2+V_l(r_n)x_n^2 \right].
\end{align}
Note that, we define $d\sum_{n=-\infty}^0\to \int_{-\infty}^{r_\ast(r_H+\epsilon)} dr_\ast$ when $d\to 0$.

The full action is a sum of the fields in the left and the right wedges. As we saw in the previous section, 
 the path integral containing both the left and right fields
can be rewritten by a path integral of a right field
on a closed time path.
In the previous section, we have
written the filed in the lower line by $\tilde{x}_R$.
In the following, 
we use a unified notation and
 write $x_R$ by $x^1$ and $\tilde{x}_R$
by $x^2$.

Previously we considered a path from $t=-\infty$
to $\infty$.
It can be generalized to a path up to a finite time
with fixed boundary conditions 
 $x_0^I(t)=x^{I}_{\text{fin}}$ ($I=1,2$);
\begin{align}
P[x^{I}_{\text{fin}},t]&=\int^{x^{I}_0(t)=x^{I}_{\text{fin}}} \prod_{n=-\infty}^{0}  {\cal D}x_{n}^{1}{\cal D}x_{n}^{2}
e^{iS[x^1_{-N},\cdots , x^1_0,x^1_1] -iS[x^{2}_{-N},\cdots , x^{2}_0,x^{2}_1]}
\label{probability}
\end{align}

\begin{figure}[ht]
\begin{center}
\begin{overpic}[scale=0.6]{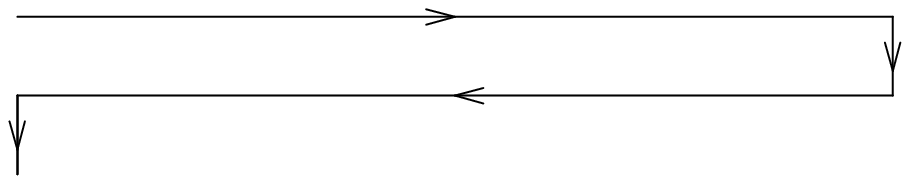}
\put(0,21){\hspace{-1.2em}$-\infty$}
\put(50,22){$x^1_n$}
\put(2,0){$-\infty-i\beta$}
\put(100,19){$t$}
\put(100,6){$t-\frac{i\beta}{2}$}
\put(65,0){$x^{2}_n$}
\end{overpic}
\end{center}
\caption{
The values of the fields at the right ends of the paths, 
$x^I_n(t)$ $I=1,2$, are fixed in the path integral. 
}
\label{path_t}
\end{figure}

By integrating the environmental variables (fields between $r_H$ and $r_H+\epsilon$),
we have
\begin{align}
P[x^{I}_{\text{fin}},t]&=\int^{x^{I}_0(t)=x^{I}_{\text{fin}}}  {\cal D}x_{0}^{1}{\cal D}x_{0}^{2}
e^{iS[x^1_0,x^1_1]-iS[x^{2}_0,x^{2}_1] +iS_{IF}[x^1_0, x^{2}_0] }.
\end{align}
The definition of the influence functional $S_{IF}$ is schematically written by
\begin{align}
e^{iS_{IF}[x^1_0, x^{2}_0] }
&\equiv
\int \prod_{n=-\infty}^{-1} {\cal D}x_{n}^{1}
{\cal D}x_{n}^{2}
e^{iS[x^1_{-N},\cdots , x^1_{-1}]
-iS[x^{2}_{-N},\cdots , x^{2}_{-1}]
+iS_{int}[x^1_{-1},x^1_0]
-iS_{int}[x^{2}_{-1}, x^{2}_0]}
,
\end{align}
where $S_{int}[x_{-1},x_0]=k/2 \int dt  (x_0-x_{-1})^2$. 

Since the {\it system} variables $x^{1}_0,x^{2}_0 $ are coupled linearly with
the environment variables, 
the influence functional $S_{IF}[x^1_0, x^{2}_0]$ 
has a Gaussian form
\begin{align}
S_{IF}[x^1_0, x^{2}_0]&=\frac{1}{2}\int dtdt' x^{I}_0(t) F_{IJ}(t,t') x^{J}_0(t').
\end{align}
The Kernel function in the Schwinger-Keldysh formalism
 $F_{IJ}(t,t')$ can be obtained 
by taking derivatives of  the influence functional as
\begin{align}
F_{IJ}(t,t') &=\frac{1}{i}\frac{\delta^2}{\delta x_0^{J}(t') \delta x_0^{I}(t)}e^{iS_{IF}[x^1_0, x^{2}_0] }|_{x_0^{I}=0}\nn
&=i (kd)^2\left(
\begin{array}{cc}
\langle T  \Delta^+ x^1_{-1}(t) \Delta^+ x^1_{-1}(t') \rangle_{\frac{\beta}{2},\frac{\beta}{2}} 
&- \langle  \Delta^+ x^1_{-1}(t) \Delta^+ x^{2}_{-1}(t') \rangle_{\frac{\beta}{2},\frac{\beta}{2}} \\
- \langle  \Delta^+ x^{2}_{-1}(t) \Delta^+ x^1_{-1}(t')\rangle_{\frac{\beta}{2},\frac{\beta}{2}} 
& \langle \tilde{T} \Delta^+ x^{2}_{-1}(t) \Delta^+ x^{2}_{-1}(t') \rangle_{\frac{\beta}{2},\frac{\beta}{2}}
\end{array}
\right),
\end{align}
The expectation means an integration over 
the environmental  variables $x^{I}_{-\infty}, \cdots , x^{I}_{-1}$ ($I=1,2$).
$T$ stands for the time ordering, and $\tilde{T}$ is the anti-time ordering. 
As we saw in the previous subsection (\ref{LR-SK}), 
these propagators are equal to the Schwinger-Keldysh (SK) ones with the path drawn in 
Fig.\ref{path_aa}. 
In the continuum limit, the discrete Green functions
$d \times F_{IJ}(t,t^\prime) $ become 
the continuum counterpart
\begin{align}
&F^{IJ}_{(l,m) (l',m')}(t,t')\nn
&=i \partial_{r_\ast}  \partial_{r_\ast '} 
\left(
\begin{array}{cc}
\langle T \phi^1_{(l,m)}(t,r_\ast) \phi^1_{(l',m')}(t',r'_\ast) \rangle_{\frac{\beta}{2},\frac{\beta}{2}} 
& -\langle \phi^1_{(l,m)}(t,r_\ast) \phi^{2}_{(l',m')}(t',r'_\ast) \rangle_{\frac{\beta}{2},\frac{\beta}{2}} \\
-\langle \phi^{2}_{(l,m)}(t,r_\ast) \phi^1_{(l',m')}(t',r'_\ast) \rangle_{\frac{\beta}{2},\frac{\beta}{2}} 
& \langle \tilde{T} \phi^{2}_{(l,m)}(t,r_\ast) 
\phi^{2}_{(l',m')}(t',r'_\ast) \rangle_{\frac{\beta}{2},\frac{\beta}{2}}
\end{array}
\right)|_{ r=r'=r_H+\epsilon}\nn
&\equiv \partial_{r_\ast}  \partial_{r_\ast '} 
\left(
\begin{array}{cc}
G^{11}_{(l,m) (l',m')}(t,r_\ast ;t',r_\ast') & G^{12}_{(l,m) (l',m')}(t,r_\ast ;t',r_\ast')\\
G^{21}_{(l,m) (l',m')}(t,r_\ast ;t',r_\ast') & G^{22}_{(l,m) (l',m')}(t,r_\ast ;t',r_\ast')
\end{array}
\right)|_{r=r'=r_H+\epsilon} 
\end{align}
and the influence functional is given by 
\begin{align}
S_{IF}[\phi^1(r_H+\epsilon), \phi^{2}(r_H+\epsilon)]
&=\frac{1}{2}\int dtdt'  
\phi^{I}_{(l,m)}(t, r_H+\epsilon)F^{IJ}_{(l,m),(l^\prime,m^\prime)}(t,t') \phi^{J}_{(l',m')}(t^\prime, r_H+\epsilon).
\end{align}
Strictly speaking, the expectation in  the Green functions
 should be evaluated at $(r_\ast)_{-1}$,
but in the continuum limit it coincides with the position at the stretched horizon
at $r_H+\epsilon$.

\subsubsection{Vacuum Condition}
From the previous discussions, we already knew that the Green
functions in the Kruskal vacuum become identical with the Schwinger-Keldysh Green functions
along the contour in Fig.\ref{path_aa}.  
We repeat the 
discussion for the case of the two point functions explicitly in the following.
In order to calculate the influence functional, we need to specify
the vacuum condition for the environmental variables,
i.e. the Kruskal vacuum condition so that
the physical quantities is regular in the Kruskal coordinates.
We expand the scalar field by $u_k^{(1)} , u_k^{(2)}$ and its complex conjugates
\begin{align}
\phi_{(l,m)}(t,r_\ast)&=\int\frac{dk}{\sqrt{4\pi\omega_k}}\left[
c_{k (l,m)}^{(1)} u_k^{(1)}+(c_{k (l,m)}^{(1)})^\dagger (u_k^{(1)})^\ast
+
c_{k (l,m)}^{(2)} u_k^{(2)}+(c_{k (l,m)}^{(2)})^\dagger (u_k^{(2)})^\ast
\right],
\end{align}
with the canonical commutation relations
\begin{align}
[c_{k (l,m)}^{(1) },(c_{k' (l',m')}^{(1) })^\dagger ]&=(2\pi)2\omega_k \delta_{l l'}\delta_{mm'}\delta(k-k'), \\
[c_k^{(2) (l,m) },(c_{k'}^{(2) (l',m')})^\dagger  ]&=(2\pi)2\omega_k \delta_{l l'}\delta_{mm'}\delta(k-k') .
\end{align}
The correlators in the Kruskal vacuum become the following forms,
\begin{align}
F^{AB}_{K, (l,m) (l',m')}(t,t') &=
\delta_{ll'}\delta_{mm'}\partial_{r_\ast}\partial_{r_\ast'} G_K^{AB}(t,r_\ast ; t^\prime,r_\ast^\prime)_{|r_\ast=r_\ast^\prime}
\end{align}
where 
\begin{align} 
G_K^{AB}(t,r_\ast ; t^\prime, r_\ast^\prime) &= i
\left( \begin{array}{cc}
\langle T \phi^R(t,r_\ast) \phi^R(t^\prime,r_\ast^\prime) 
\rangle_K &
\langle \phi^R(t,r_\ast) \phi^L(t^\prime,r_\ast^\prime) \rangle_K \\
\langle \phi^L(t,r_\ast) \phi^R(t^\prime,r_\ast^\prime) \rangle_K &
\langle T \phi^L(t,r_\ast) \phi^L(t^\prime,r_\ast^\prime) \rangle_K
\end{array} \right) 
\label{RL}
\end{align}
Here $K$ means the expectation value in the Kruskal vacuum.
As we saw, they are related to the
Schwinger-Keldysh Green functions $F^{IJ}_{(l,m), (l',m')}(t,t')=\delta_{ll'}\delta_{mm'}\partial_{r_\ast}\partial_{r'_\ast}G^{IJ}(t,t')|_{r=r'=r_H+\epsilon}$
discussed in the previous section as
\begin{align}
&\frac{1}{2}\int dt dt' \phi^A_{(l,m)}(t,r_H+\epsilon)
F^{AB}_{K, (l,m), (l',m')}(t,t')
\phi^B_{(l',m')}(t',r_H+\epsilon) \nn
& \ =\frac{1}{2}\int dt dt' 
\phi^{I}_{(l,m)}(t,r_H+\epsilon)
F^{ IJ}_{(l,m), (l',m')}(t,t')
\phi^{I}_{(l',m')}(t',r_H+\epsilon),
\end{align}
where the Schwinger-Keldysh Green functions 
$G^{IJ}$ are given by 
\begin{align} 
G^{IJ}(t,r_\ast ; t^\prime, r_\ast^\prime) &= i
\left( \begin{array}{cc}
\langle T \phi^1(t,r_\ast) \phi^1(t^\prime,r_\ast^\prime) 
\rangle_{\frac{\beta}{2},\frac{\beta}{2}} &
- \langle \phi^1(t,r_\ast) \phi^{2}(t^\prime,r_\ast^\prime) 
\rangle_{\frac{\beta}{2},\frac{\beta}{2}} \\
- \langle \phi^{2}(t,r_\ast) \phi^1(t^\prime,r_\ast^\prime) 
\rangle_{\frac{\beta}{2},\frac{\beta}{2}} &
\langle \tilde{T} \phi^{2}(t,r_\ast) \phi^{2}
(t^\prime,r_\ast^\prime) \rangle_{\frac{\beta}{2},\frac{\beta}{2}}
\end{array} \right) \nn
&=i \int \frac{dk}{4\pi \omega_k} 
 \frac{1}{2 \sinh(\pi \omega_k/\kappa)} 
 \left(
\begin{array}{cc}
M^{11}(t,t^\prime)
 & M^{12}(t,t^\prime)
\\
M^{21} (t,t^\prime)
&
M^{22}(t,t^\prime)
\end{array} \right)
e^{ik(r_\ast-r_\ast^\prime)} .
\end{align}
Non-diagonal entries have an extra minus sign with respect to eq.(\ref{RL}), 
since the $\phi^2$ field is defined to evolve backward
in time as in Fig.\ref{path_aa}.

Each component can be calculated as
\begin{align}
M^{11}(t,t^\prime) &=\theta(t-t^\prime)\left(  e^{\frac{\pi \omega}{\kappa}} e^{-i\omega (t-t^\prime)}
+ e^{-\frac{\pi \omega}{\kappa}} e^{i\omega (t-t^\prime)}  \right) \nn
&\hspace{2em}+\theta(t^\prime-t) \left(  e^{-\frac{\pi \omega}{\kappa}} e^{-i\omega (t-t^\prime) }
+ e^{\frac{\pi \omega}{\kappa}} e^{i\omega (t-t^\prime)}
\right),
\\
M^{22}(t,t^\prime)& =\theta(t^\prime-t)\left(  e^{\frac{\pi \omega}{\kappa}} e^{-i\omega (t-t^\prime)}
+ e^{-\frac{\pi \omega}{\kappa}} e^{i\omega (t-t^\prime)}
\right) \nn
&\hspace{2em}+\theta(t-t^\prime) \left(  e^{-\frac{\pi \omega}{\kappa}} e^{-i\omega(t-t^\prime)  }
+ e^{\frac{\pi \omega}{\kappa}} e^{i\omega (t-t^\prime)}
\right), \nn
M^{12}(t,t^\prime) &=M^{21}(t,t^\prime) =e^{-i\omega (t-t^\prime)}+ e^{i\omega (t-t^\prime)}.
\end{align}
It can be  also rewritten in the following form,
\begin{align}
G^{IJ}(t,r_\ast; t^\prime,r_\ast^\prime) &=
\int\frac{dk_0 dk}{(2\pi)^2} e^{-ik_0 (t-t')+ik(r_\ast - r_\ast')}
\nn
&
\left(
\begin{array}{cc}
\frac{1}{-k_0^2+\omega_k^2-i\epsilon}+2\pi i n(\omega_k)\delta(-k_0^2+\omega_k^2) & 
-2\pi i \sqrt{n(1+n)}\delta(-k_0^2+\omega_k^2) \\
-2\pi i \sqrt{n(1+n)}\delta(-k_0^2+\omega_k^2) &
\frac{-1}{-k_0^2+\omega_k^2+i\epsilon}+2\pi i n(\omega_k)\delta(-k_0^2+\omega_k^2) 
\end{array}
\right) ,
\end{align}
where $n(\omega_k)=1/(e^{\beta_H \omega_k}-1)$, $\beta_H=2\pi/\kappa$. 
The 2-2 component of the Green function coincides with the 
anti-time ordered finite temperature Green function, while the 
1-1 component
is the ordinary time ordered one. 
\\

\begin{figure}[ht]
\begin{center}
\begin{overpic}[scale=0.6]{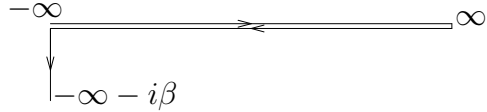}
\put(0,21){\hspace{-1.2em}$-\infty$}
\put(2,0){$-\infty-i\beta$}
\put(100,19){$\infty$}
\end{overpic}
\caption{
This path corresponds to the propagators which have non-diagonal entries (\ref{SKcorrelator}).}
\label{path_b}
\end{center}
\end{figure}

In the conventional real-time finite-temperature field theory, 
the contour is usually taken as in the figure \ref{path_b}. 
The contour corresponds to considering an ordinary finite temperature
Green function with the Boltzmann factor $e^{-\beta H}$ at the left-end;
\begin{align}
\langle \hat{\cal O}^1 
\hat{\cal O}^2 \rangle_\beta
&= \frac{1}{Z} 
\text{Tr} \left( e^{-\beta H} 
\hat{\cal O}^1(t) \ \hat{\cal O}^2(t')
\right)
\end{align}
irrespective of whether these operators live 
on the upper or lower lines.
On the other hand, the Green function (\ref{LR-SK}) we are considering 
corresponds 
to taking a different contour as drawn in Fig. \ref{path_aa}.
The fields on the lower line 
in these two contours are related by the similarity transformation
\begin{align}
\phi_{\text{New}}^2(t,x)  = e^{-\beta H_R /2} \phi^2 (t,x) e^{\beta H_R/2}  
= \phi^2(t+i\beta/2).
\end{align}
Here $\phi^2$ and $\phi_{\text{New}}^2$ are fields 
appearing in the formalism of Fig \ref{path_aa}
and Fig \ref{path_b} respectively.
In the momentum representation, it is
\begin{align}
 \phi^{2}_{\text{New}}(k) = e^{\frac{\beta k_0}{2}}\phi^{2}(k) ,
\label{newfield} 
\end{align}

In terms of the new field,
the Green function  can be written in the following form,
\begin{align}
&G_{\text{New}}^{IJ}(t,r_\ast; t^\prime, r_\ast^\prime)\nn
&= i \left( \begin{array}{cc}
\langle T \phi^1(t) \phi^1(t^\prime) \rangle_\beta &
-\langle \phi^1(t) \phi_{\text{New}}^{2}(t') \rangle_\beta \\
-\langle \phi_{\text{New}}^{2}(t) \phi^1(t') \rangle_\beta &
\langle \tilde{T} \phi_{\text{New}}^{2}(t) 
\phi_{\text{New}}^{2}(t') \rangle_\beta
\end{array} \right) \nn 
&= \int\frac{dk_0 dk}{(2\pi)^2} e^{-ik_0 (t-t')+ik(r_\ast - r_\ast')}
\nn
&
\left(
\begin{array}{cc}
\frac{1}{-k_0^2+\omega_k^2-i\epsilon}+2\pi i n(\omega_k)\delta(-k_0^2+\omega_k^2) & 
-2\pi i\ \text{sgn}(k_0)n(k_0)\delta(-k_0^2+\omega_k^2) \\
-2\pi i\ \text{sgn}(k_0)(1+n(k_0))\delta(-k_0^2+\omega_k^2) &
\frac{-1}{-k_0^2+\omega_k^2+i\epsilon}+2\pi i n(\omega_k)\delta(-k_0^2+\omega_k^2) 
\end{array}
\right) .
\label{SKcorrelator}
\end{align}

\subsubsection{Langevin equation at Stretched Horizon \label{sec-Lan}}
The effective equation of motion at the stretched horizon can be
obtained by taking a variation of the effective action
 $S[x^1_0]-S[x^2_0] +S_{IF}[x^1_0, x^2_0] $.
In taking a continuum limit, a care should be taken since we have already 
integrated out the environmental field $x_{-1}$, and
only the interaction with the outer variable $x_1$ appears in the 
effective action for $x_0$.
The equation of motion for $x_0^I$ becomes
\begin{align}
\ddot{x}_0^{ I} = -k (x_0^{I}-x_1^{ I}) -  \int dt' F^{IJ} (t,t') x_0^{ J} (t').
\end{align}
In the continuum limit ($d \rightarrow 0$) with $kd^2=$fixed, 
the time derivative term drops and we have
\begin{align}
 \partial_{r_\ast}\phi^{I}_{(l,m)}(t)-\int dt' F^{ IJ}_{(l,m) (l',m')}(t,t')\phi^{J}_{(l',m')}(t')=0.
\end{align}
(Note that the discretized $d \times F^{ IJ}$ becomes the continuum $F^{ IJ}_{(l,m) (l',m')}$.)
The dynamics seems to have disappeared in the effective equation at the stretched
horizon, but we will see that another time derivative term (which is first order)
is induced from the second term. 

In order to show this, following the retarded-advanced formalism
discussed below, we recombine the Schwinger-Keldysh fields, 
 $\phi^1(x)$ and $\phi^{2}(x)$,
into  a {\it classical} variable $\phi^r(x)$ 
and a {\it fluctuating} variable $\phi^a(x)$. 
The interpretation of {\it classical} and {\it fluctuating} variables comes from 
an observation that
 the action $S[\phi^1]-S[\phi^2]$ 
has a dominant contribution in the path integral when
the configuration of two fields coincide. 
As we saw in Fig. \ref{path_aa},  
the time axis of $\phi^1(t)$ differs from that of $\phi^2(t)$ by 
an amount of $\beta/2$ into the imaginary direction,
and
dominant configurations are given in terms of the redefined 
field (\ref{newfield})  $\phi^2_{\text{New}}(t) = \phi^1(t+i \beta/2)$ in the following way.
We define the {\it classical} and {\it fluctuating} fields as
\begin{align}
\left\{
\begin{array}{l}
\phi^r_{(l,m)} = \frac{1}{\sqrt{2}}\left(
\phi^1_{(l,m)}  +\phi^{2}_{\text{New} (l,m)}  \right)\\
\phi^a_{(l,m)}= \frac{1}{\sqrt{2}}\left( \phi^1_{(l,m)} 
-\phi^{2}_{\text{New} (l,m)} \right). 
\end{array}
\right.
\end{align}
Propagators are transformed in this basis as 
\begin{align}
&
\left( \phi^1\ \phi^{2}_{\text{New}} \right)
\left(
\begin{array}{cc}
G_{\text{New}}^{11} & G_{\text{New}}^{12}\\
 G_{\text{New}}^{21} & G_{\text{New}}^{22}
\end{array}
\right)
\left(
\begin{array}{c}
\phi^1\\
\phi^{2}_{\text{New}}
\end{array}
\right)
=
\left( \phi^r\ \phi^a \right)
\left(
\begin{array}{cc}
0 & G^A \\
G^R & 2iG^{\text{sym}}
\end{array}
\right)
\left(
\begin{array}{c}
\phi^r\\
\phi^a
\end{array}
\right).
\end{align}
where we have defined
\begin{align}
G^R(t)&=\frac{1}{2}(G_{\text{New}}^{11}-G_{\text{New}}^{12}
+G_{\text{New}}^{21}-G_{\text{New}}^{22})(t) 
=i \theta(t)\langle [\phi(t),\phi(0)]\rangle\\
G^A(t)&=\frac{1}{2}(G_{\text{New}}^{11}
+G_{\text{New}}^{12}-G_{\text{New}}^{21}
-G_{\text{New}}^{22})(t)=-i \theta(-t)\langle [\phi(t),\phi(0)]\rangle \\
G^{\text{sym}}(t)&=-\frac{i}{4}(G_{\text{New}}^{11}
-G_{\text{New}}^{12} -G_{\text{New}}^{21}+G_{\text{New}}^{22})(t)
=\frac{1}{2}\langle \{\phi(t), \phi(0) \}\rangle
\end{align}
and  used the relation
$
G_{\text{New}}^{11}+G_{\text{New}}^{12}
+G_{\text{New}}^{21}+G_{\text{New}}^{22}=0.
$
Because of this, the basis $\left( \phi^r\ \phi^a \right)$ are often called 
the retarded-advanced basis.

In terms of the $r,a$-fields 
the influence functional can be written as
\begin{align}
S_{IF}&=\int dtdt'  \Big[\phi^a_{(l,m)}(t)\partial_{r_\ast}\partial_{r_\ast'}G^R_{(l,m) (l',m')}(t,t')\phi^r_{(l',m')}(t') \nn
&\hspace{6em} +i \phi^a_{(l,m)}(t)\partial_{r_\ast}\partial_{r_\ast'}G^{\text{sym}}_{(l,m) (l',m')}(t,t')\phi^a_{(l',m')}(t')\Big].
\end{align}
The derivative of the retarded Green function  $\partial_{r_\ast}\partial_{r_\ast '}G^R$
satisfies
\begin{align}
\partial_{r_\ast}\partial_{r_\ast '}G^R_{(l,m) (l',m')}(t,t')|_{r=r'=r_H+\epsilon} &=
-\delta_{l l'}\delta_{m m'}\partial_{t'}\delta(t-t'),
\end{align} 
On the other hand, the symmetric Green function can be written 
as 
\begin{align}
G^{\text{sym}}=\int \frac{dk}{4\pi \omega_k}\left(n+\frac{1}{2}\right)
\left( e^{-i\omega_k (t-t^\prime)}+e^{+i\omega_k (t-t^\prime)}\right)
e^{ik(r_\ast-r_\ast^\prime)},
\end{align}
and its derivative becomes 
\begin{align}
\partial_{r_\ast}\partial_{r_\ast '}G^{\text{sym}}_{(l,m) (l',m')}(t,t')|_{r=r'=r_H+\epsilon}
&= \delta_{l l'}\delta_{m m'} \int \frac{dk_0}{4\pi} \frac{k_0}{\tanh \frac{\beta k_0}{2}}e^{-ik_0(t-t')} 
\nn
&= \delta_{l l'}\delta_{m m'}\frac{1}{2\pi}\left[-\frac{\kappa^2}{4 \sinh^2\frac{\kappa (t-t')}{2}} \right].
\end{align}
for $t\neq t'$.
The integral is divergent at $t=t^\prime$. 
Since we are interested in the
finite temperature effect, we regularize the symmetrized correlator
by removing the $\kappa$-independent
divergence (note $T=\kappa/2\pi$) as
\begin{align}
K(t,t') \equiv
\frac{1}{2\pi}\left[-\frac{\kappa^2}{4 \sinh^2\frac{\kappa (t-t')}{2}} +\frac{1}{(t-t')^2}\right].
\end{align}
Hence the  action for the stretched horizon variable,
which is a sum of $S_{IF}$ and the interaction term with
the neighboring variable $x_1$, becomes 
\begin{align}
S &= \int dt dt' d\Omega r_\epsilon^2 \Big[ 
\phi^a(t,r_\epsilon,\Omega)\delta(t-t') (\partial_{t'}-\partial_{r^\ast}) \phi^r(t',r_\epsilon,\Omega)\nn
&\hspace{10em}+ i \phi^a(t,r_\epsilon,\Omega) K(t,t') \phi^a(t',r_\epsilon,\Omega) \Big].
\label{eff_action}
\end{align}
Here, $r_\epsilon\equiv r_H+\epsilon$ appears with rewriting $\phi^r_{(l,m)}$ to $\phi^r$.
By integrating the fluctuating variable $\phi^a(t)$, 
(\ref{probability})
is written as
\begin{align}
&P[\phi^r_{\text{fin}},t] =\nn
& \int^{\phi^r(t)=\phi^r_{\text{fin}}} {\cal D} \phi^r 
\exp \left[
-\frac{1}{4} \int dt dt' d\Omega r_\epsilon^2(\partial_{t}-\partial_{r^\ast}) \phi^r(t,r_\epsilon,\Omega) K^{-1}(t,t') 
(\partial_{t'}-\partial_{r'^\ast}) \phi^r(t',r_\epsilon,\Omega)\right] 
\label{EA-SH}
\end{align}
It describes the effective dynamics at the stretched horizon.
Note that the effective action contains a term which is odd 
under the time reversal transformation.

Instead of integrating out the fluctuating variable,
we can  introduce an auxiliary field $\xi(t)$ by
\begin{align}
& \exp \left( 
-\int dt dt'\phi^a_{(l,m)}(t)
K(t,t') \phi^a_{(l,m)}(t')
\right) \nonumber \\
&=\int{\cal D}\xi \exp \left(
i\int dt \phi^a_{(l,m)}(t) \sqrt{2}\xi_{(l,m)}(t)
-\frac{1}{2}\int dt dt' \xi_{(l,m)}(t) K^{-1}(t,t') \xi_{(l,m)}(t') 
\right).
\end{align}
Then the probability to see $\phi^r(t)=\phi^r_{\text{fin}}$ at the stretched horizon is written 
in terms of the scalar fields  $\phi^{r,a}(t)$ and 
the auxiliary field $\xi$ as
\begin{align}
P[\phi^r_{\text{fin}},t] &=\int^{\phi^r(t)=\phi^r_{\text{fin}}} {\cal D}\phi^r  {\cal D}\phi^a{\cal D}\xi\
e^{-\frac{1}{2}\int dt dt' \xi_{(l,m)}(t) K^{-1}(t,t')
\xi_{(l',m')}(t') }
e^{iS_{\text{eff}}[\phi(t),\xi]},\\
S_{\text{eff}}[\phi(t),\xi]
&=\int  dt \phi^a_{(l,m)}(t)\left[ -\partial_{r_\ast}\phi^r_{(l,m)}(t) +\int dt' G^R_{(l,m) (l',m')}(t,t')\phi^r_{(l',m')} (t')+\sqrt{2}\xi_{(l,m)}(t)\right] .
\end{align}
The variation with respect to $\phi^a$
gives 
the equation of motion for $\phi^r$
\begin{align}
(\partial_t-\partial_{r_\ast})\phi^r_{(l,m)}+\sqrt{2}\xi_{(l,m)}(t)&=0,
\label{Original_Langevin}
\end{align}
with the Gaussian noise correlation 
\begin{align}
\langle \xi_{(l,m)}(t) \rangle&=0 \ , \ 
\langle \xi_{(l,m)}(t)\xi_{(l' ,m')}(t') \rangle=\delta_{l l'}\delta_{m m'}
K(t,t').
\label{noise}
\end{align}
As expected, if we take the statistical average,
the outgoing modes vanish in the averaged sense $\langle (\partial_t-\partial_{r_\ast})\phi^r \rangle =0$,
which means that 
there are only ingoing modes at the (stretched)
horizon. The noise term 
can be considered as the effect of the 
Hawking radiation.
In the next subsection, we compare
the noise correlation obtained here 
with the flux of the Hawking radiation.
The noise correlation is not white, and 
the memory effect remains with a time scale of the
Hawking temperature $(t-t^\prime) \sim 1/\kappa = \hbar/2\pi T_H$.
If we look at the dynamics of a time scale larger than it,
we can approximate the noise as the following white noise
\begin{align}
\langle \xi_{(l,m)}(t)\xi_{(l' ,m')}(t') \rangle
\longrightarrow \delta_{l l'} \delta_{m,m^\prime} \frac{\kappa}{2\pi}
\delta(t-t^\prime)
=\delta_{l l'} \delta_{m,m^\prime} T_H
\delta(t-t^\prime)
\label{whitenoise}
\end{align}

The above effective action is  obtained previously
based on the physical picture of the Hawking radiation \cite{deBoer:2008gu}
or a technique to reproduce the Schwinger-Keldysh formalism \cite{Son:2009vu}
in a setting of vibrating string in AdS black hole background.
We have reproduced the same effective action by
explicitly integrating the environmental variables
between the horizon and the stretched horizon.
Because of the mixing of the wave functions (\ref{Unruh}),
the integration corresponds to an integration over the
variables hidden in the horizon.

\section{The Fluctuation Theorem for Black Holes and Matters\label{sec-FTBH}}
Now we apply the fluctuation theorem to the scalar field 
in the black hole background. 
Most generally, we must treat the whole system 
of the scalar field and the space-time
as a coupled quantum system, and backreactions to the 
space-time structure
must be included.
In our previous letter \cite{Iso:2010tz}, we briefly sketched how to
treat the metric degrees of freedom quantum mechanically
in the path integral formalism and discussed the 
effect of the backreaction. 
In the present paper, in order to 
give a more systematic and complete investigation,
we  consider an easier situation, i.e.
a scalar system in a fixed black hole background.
We neglect  effects of backreactions. 
Even if we adopt such a simplification, 
various interesting results  follow
 the fluctuation theorem applied to our system,
such as a proof of the generalized second law
and  a derivation of the Green-Kubo formula.

\subsection{Discretized Equations outside the Stretched Horizon}
The equation of motion of 
the scalar field $\phi^r(t,r_\ast)$ in the black hole background consists
of the two coupled equations, namely, the effective equation 
at the stretched horizon $r=r_H+\epsilon$ and the bulk equation of motion
outside the stretched horizon.

We put the scalar field in a box with a radius $r_B (> r_H)$ and 
impose a boundary condition at the outer boundary 
$\phi(t,r=r_B, \Omega)=0$ in this subsection.
Owing to the boundary condition, the scalar field 
is shown to be thermalized.
Another merit of confining the system in a box is to 
stabilize the total system (even we take the backreaction 
into account \cite{Gibbons:1976pt} if the size of the box is 
not so large.)
It thus justifies to choose an equilibrium distribution as an initial distribution for the matter field as we will do in the following.
In a later section, we choose a different boundary condition
to realize a steady state with a constant energy flux.

In  order to apply the fluctuation theorem 
reviewed in section \ref{sec-FT}, we need to construct a
Fokker-Planck equation which is local in time.
In doing so, it is necessary to approximate the noise 
correlation (\ref{noise}) by the white noise (\ref{whitenoise}).
This approximation is valid  for a longer time scale than
$\hbar/2\pi T_H$. Though the validity is limited, 
 we consider such an approximation in the present paper.
The memory effect of the colored noise will be discussed
later.

In the white noise approximation, the discretized equations
are given by
\begin{align}
\left\{
\begin{array}{l}
\gamma_0 \dot{x}_0=-k(x_0-x_1) - \sqrt{2}\xi_0 \ , \ \ 
 \langle\xi_0(t)\xi_0(t')\rangle=
\gamma_0 T_H 
\delta (t-t')\\
\ddot{x}_1 =-k(x_1-x_2)+k(x_0-x_1)- V_l(r_1)x_1\\
\vdots\\
\ddot{x}_N =-k(x_N-x_{N+1})+k(x_{N-1}-x_N)- V_l(r_N)x_N \\
x_{N+1}\equiv 0.
\label{discretized}
\end{array}
\right.
\end{align}

\begin{figure}[ht]
\begin{center}
\begin{overpic}[scale=0.6]{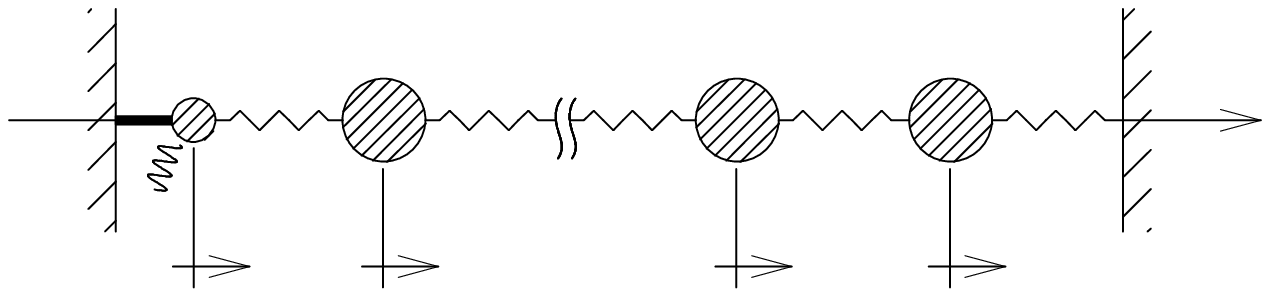}
\put(10,4){$\xi_0$}
\put(10,16.5){$\gamma_0$}
\put(21,16){$k$}
\put(38,16){$k$}
\put(49,16){$k$}
\put(65,16){$k$}
\put(82,16){$k$}
\put(16,4){$x_0$}
\put(32,4){$x_1$}
\put(59,4){$x_{N-1}$}
\put(76,4){$x_{N}$}
\put(100,13){$r_\ast$}
\end{overpic}
\caption{A schematic illustration of eq.(\ref{discretized}).
Each variables $x_i$ are bound by spring with constant $k$.
Only $x_0$ reserves a friction $\gamma_0$ and a noise $\xi_0$.
The effects of $V_l(r_i)$ are not described on picture,
it affects each variables as harmonic potential.
}
\end{center}
\label{model}
\end{figure}

The first line is the stochastic equation for the field at the inner boundary
 (stretched horizon) $r= r_\epsilon \equiv r_H +\epsilon$  with a noise term $\xi$.
Note that the time derivative term originates in the dissipation
term induced by the interaction with the environmental variables. 
The last one is the boundary condition at the outer boundary $r=r_B$.
The middle ones are the bulk equations of motion, and
the field $\phi_{l,m}(r_\ast)$ between the stretched horizon and
the outer boundary is discretized into $N$ lattice points in 
the tortoise coordinate $r_\ast$.
The continuum limit can be taken as before by
taking  $d \to 0$ (or $N \to \infty$)
with the following conditions and replacements;
\begin{align}
(N+1)d=r_\ast(r_B)-r_\ast(r_\epsilon), \ 
kd^2 \equiv 1 \ , \ \gamma_0 d\equiv 1   ,\ \phi^r_{(l,m)}((r_\ast)_i)\equiv x_i/\sqrt{d}, \ \xi_{(l,m)}\equiv \sqrt{d}\xi_0,
\end{align}
where $d$ is the lattice spacing in the tortoise coordinate $r_\ast$.
The continuum equations in the bulk can be recovered as before and becomes
\begin{align}
 \ddot{\phi}^r_{(l,m)}(t,r_\ast)&=\partial_{r_\ast}^2\phi^r_{(l,m)}(t,r_\ast)-V_l(r_\ast)\phi^r_{(l,m)}(t,r_\ast). 
\end{align}

At the stretched horizon, 
the first equation of (\ref{discretized}) can be written as
\begin{align}
\gamma_0  \dot{x}_0
&=kd \Delta^+ x_0 -\sqrt{2}\xi_0 \nn
  \xrightarrow{d\to 0}  \dot{\phi}^r_{(l,m)}(t, r_\epsilon)&=\partial_{r_\ast} 
\phi^r_{(l,m)}(t,r_\epsilon) -\sqrt{2} \xi_{(l,m)},
\end{align}
with noise correlation
\begin{align}
\langle \xi_{(l,m)}(t) \xi_{(l',m')}(t') \rangle&=\delta_{ll'}\delta_{mm'}
d\langle \xi_0(t) \xi_0(t') \rangle=
\delta_{ll'}\delta_{mm'}d \gamma_0 T_H\delta(t-t')\nn
&=\delta_{ll'}\delta_{mm'} T_H\delta(t-t').
\end{align}

\subsection{Fluctuation Theorem for Scalar Field in Black Hole Background}
From the set of the Langevin equations
(\ref{discretized}),
we can construct 
 the corresponding 
Fokker-Planck equation of $P(x_0,x_1,\cdots ,x_N , v_1 , \cdots , v_N ,t)$
with $2N+1$ set of variables;
\begin{align}
\partial_t P&=\partial_{x_0}\left[\frac{1}{\gamma_0} \frac{\partial U}{\partial x_0}P+\frac{T_H}{\gamma_0}\partial_{x_0}P \right]
+\sum_{i=1}^N\left[\partial_{x_i}(-v_i P)+\partial_{v_i}\left(
 \frac{\partial U}{\partial x_i}P\right)\right],
\end{align}
where we 
defined $U(x)\equiv \frac{1}{2}\sum_{i=1}^{N+1}\left[(\Delta^- x_i)^2+V_l(r_i)x_i^2 \right]$.
In this expression, we introduced a redundant variable
$x_{N+1}$ for convenience, but  eventually set $x_{N+1}=0$.
This equation has a solution describing an equilibrium distribution, 
\begin{align}
& P^{\text{eq}}=Z^{-1} e^{-\frac{1}{T_H} \left[ \frac{1}{2}\sum_{i=1}^Nv_i^2 +U(x)\right]}, 
\nn
& Z=\int d^{N+1}x d^N v e^{-\frac{1}{T_H} \left[ \frac{1}{2}\sum_{i=1}^Nv_i^2 +U(x)\right]}.
\end{align}
A general solution to the Fokker-Planck equation can be formally 
represented in the path integral form as, 
\begin{align}
 P(x_0 , x_i , \tau|x'_0 , x'_i ,t=0 )\propto \int_{x'}^x \prod_{k=0}^N{\cal D}x_k 
e^{-\frac{1}{4\gamma_0 T_H}\int_{\Gamma_\tau} dt (\gamma_0\dot{x}_0-k(x_1-x_0))^2}
\prod_t \prod_{i=1}^N\delta(\ddot{x}_i+\tfrac{\partial U}{\partial x_i}) |_{x_{N+1}\equiv 0}.
\end{align}

The probability that a trajectory $\Gamma_\tau=\{(x'_0,x'_1,\cdots ,x'_N)\to (x_0,x_1,\cdots ,x_N)\}$ is realized is given by
\begin{align}
P[\Gamma_\tau|x']\propto 
e^{-\frac{1}{4\gamma_0 T_H}\int_{\Gamma_\tau} dt (\gamma_0\dot{x}_0-k(x_1-x_0))^2}
\prod_t \prod_{i=1}^N\delta(\ddot{x}_i+\tfrac{\partial U}{\partial x_i}).
\end{align}
On the other hand,  the probability that the reversed trajectory 
$\Gamma^\ast_\tau=\{(x_0,x_1,\cdots ,x_N) \to (x'_0,x'_1,\cdots ,x'_N)\}$ 
is realized is given by 
\begin{align}
P[\Gamma^\ast_\tau|x]\propto 
e^{-\frac{1}{4\gamma_0 T_H}\int_{\Gamma_\tau} dt (-\gamma_0\dot{x}_0-k(x_1-x_0))^2}
\prod_t \prod_{i=1}^N\delta(\ddot{x}_i+\tfrac{\partial U}{\partial x_i}).
\end{align}
Hence the ratio of these two probabilities 
becomes 
\begin{align}
\frac{P_{(l,m)}[\Gamma_\tau|x']}{P_{(l,m)}[\Gamma^\ast_\tau|x]}&=
\exp\left[\frac{1}{T_H}\int_{\Gamma_\tau} dt \dot{x}_0 k(x_1-x_0)\right] 
\nn
&= \exp\left[\frac{1}{T_H} \int_{\Gamma_\tau} dt 
 \dot{\phi}^r_{(l,m)}(r_\epsilon) 
\partial_{r_\ast}\phi^r_{(l,m)}(r_\epsilon)\right].
\end{align}
Here we have written the index for the angular momentum $(l,m)$
explicitly. 
Summing over all the contributions from
various partial waves $(l,m)$, the ratio
can be written as an integral over the stretched horizon;
\begin{align}
\prod_{(l,m)} \frac{P_{(l,m)}[\Gamma_\tau|x']}
{P_{(l,m)}[\Gamma^\ast_\tau|x]}
&= \exp\left[\frac{1}{T_H} \int_{\Gamma_\tau} dt d\Omega \ r_\epsilon^2
 \dot{\phi}^r(t,r_\epsilon,\Omega) \partial_{r_\ast}\phi^r(t,r_\epsilon,\Omega)\right] \nn
&=\exp\left[\frac{1}{T_H} \int_{\Gamma_\tau} dt d\Omega \  r_\epsilon^2
 T^r_t(t,r_\epsilon,\Omega)\right].
\label{ratio_of_trajectory}
\end{align}
Here we have used the definition of the energy-momentum tensor
$T^r_t=\partial_t\phi^r \partial^r\phi^r
=\partial_t\phi^r \partial_{r_\ast}\phi^r$.
Logarithm of the ratio is proportional to the energy flux 
into the black hole 
$\Delta M[\Gamma_\tau]=\int_{\Gamma_\tau} dt d\Omega \  r_\epsilon^2 T^r_t(t,r_\epsilon,\Omega)$.
Hence,
 by using the first law of black hole thermodynamics $T_H \Delta S_{BH}[\Gamma_\tau]=\Delta M[\Gamma_\tau]$,
we can interpret this entropy production 
as an amount of difference of the black hole entropy during $t=0\sim\tau$,
\begin{align}
\frac{P[\Gamma_\tau|x']}{P[\Gamma^\ast_\tau|x]}&=
\exp\left[\Delta S_{BH}[\Gamma_\tau]\right].
\end{align}

In a more general setting, we can introduce an 
externally controlled parameter such as a
variable mass term $m(t)$ in the 
potential $U(x; \lambda_t^F)$.
Even in the presence of such an external parameter, 
the ratio can be shown to be  given by the difference of the 
entropy,
\begin{align}
\frac{P^F[\Gamma_\tau|x']}{P^R[\Gamma^\ast_\tau|x]}&=
\exp\left[\Delta S_{BH}[\Gamma_\tau]\right].
\end{align} 
In a case with  time-dependent external parameters, 
the forward and the reversed protocols are generally 
different and
we need to put $F$ and $R$ to distinguish them.

In order to apply the fluctuation theorem, we further multiply
the above probabilities
$P^F[\Gamma_\tau|x']$ (or $P^R[\Gamma^\ast_\tau|x]$)
by  probabilities for  the initial distributions. 
As we discussed above we can assume that the system is in an 
equilibrium distribution at the external parameter $\lambda^F_0$
 (or $\lambda^F_\tau$)
with the Hawking temperature $P^{\text{eq}}(x' ;\lambda^F_0)$
(or $P^{\text{eq}}(x; \lambda^F_\tau)$).
Hence
\begin{align}
&\frac{P^F[\Gamma_\tau|x'] P^{\text{eq}}(x' ;\lambda^F_0)}{P^R[\Gamma^\ast_\tau|x] P^{\text{eq}}(x; \lambda^F_\tau)}\nn
&=
\exp\left[\Delta S_{BH}[\Gamma_\tau] -\beta\left(H[x'; \lambda^F_0]-H[x; \lambda^F_\tau] \right)+\beta\left(F(\lambda^F_0)-F(\lambda^F_\tau) \right)\right]\nn
&=\exp\left[(\Delta S_{BH} +\Delta S_M)[\Gamma_\tau] \right].
\label{ratio}
\end{align}
Here, we defined 
the entropy difference of the matter by
$\Delta S_M =
-\beta(H[x'; \lambda^F_0]-H[x; \lambda^F_\tau] )+\beta (F(\lambda^F_0)-F(\lambda^F_\tau) )$,
where $H[x'; \lambda^F_0]$ is the total energy of the system at $t=0$ with 
an external parameter $\lambda^F_0$ and
$F(\lambda^F_0)$ is the free energy defined by $Z(\lambda^F_0)=e^{-\beta F(\lambda^F_0)}$.

The fluctuation theorem  is a direct consequence of the 
above key relation (\ref{ratio}).
As we saw in sec.\ref{sec-FT}, it is straightforward to prove that
\begin{align}
\frac{\rho^F(\Delta S_{BH}+\Delta S_{M})}{\rho^R(-(\Delta S_{BH}+\Delta S_{M}))}&=e^{\Delta S_{BH}+\Delta S_{M}}.
\label{BHFT}
\end{align}
Here 
$\rho^F(\Delta S_{BH}+\Delta S_{M})$ is the probability 
to observe a  value of the total entropy production 
$\Delta S_{BH}+\Delta S_{M}$
with the
forwardly controlled external parameter. 
The denominator is similarly defined as the probability to
observe a negative value of the entropy production
in the reversed protocol. 
Since the right hand side is usually much bigger than 1, 
the numerator is generally much bigger than the denominator.

By integrating it, we have the Jarzynski equality;
\begin{align}
\langle e^{-(\Delta S_{BH}+\Delta S_M)}\rangle =1.
\label{BHJar}
\end{align}
We observe that there must exist a path with
 $(\Delta S_{BH}+\Delta S_{M})<0$, i.e.
an entropy decreasing path,
otherwise the Jarzynski equality cannot be satisfied.
As we saw in section \ref{sec-FT}, the generalized second law \cite{Bekenstein:1974ax}
\begin{align}
\langle (\Delta S_{BH}+\Delta S_M)\rangle \geq 0.
\end{align}
is derived using
 the Jensen inequality $\langle e^x \rangle \geq e^{\langle x \rangle}$.
The above theorems  (\ref{BHFT}) and (\ref{BHJar})
are also applicable to dynamical processes which are
generally in non-equilibrium distributions,
if the Fokker-Planck equation we have used is valid.
As we noticed, the validity holds when the time scale 
of the dynamics is longer than the time scale of the 
inverse Hawking temperature
$\hbar/(2\pi T_H)$.
The condition is not always satisfied, and in such  situations, 
we need to take  effects of time-correlations of emissions.

\subsection{Memory Effect and  Quantum Corrections \label{Sec-Memory}}
In the previous sections, we have approximated the dynamics
of the scalar fields by Langevin and Fokker-Planck equations.
The approximation is valid when 
 the noise correlation 
(\ref{noise}) can be replaced by the white noise
and also the evolution of the scalar field $\phi^r$ is dominated
by the classical path described by the Langevin equation. 
The first condition is violated for a shorter time scale than
$\hbar/2\pi T_H$. The second condition is related to a
justification of the Markovian process we have used.
If we take $\hbar \to 0$ limit while
keeping $T_H= \hbar \kappa/2\pi$ fixed,  both conditions are satisfied.
If these conditions are violated, we need to treat the system
quantum mechanically without using the classical stochastic 
equations. 
More detailed analysis of such quantum corrections
will be reported in a separate paper, and a brief sketch is given here.

We start from the action (\ref{eff_action}) 
at the stretched horizon.
Before integrating out the variable $\phi^a$,
this gives an amplitude of the stretched horizon variables
$\phi^1$ and $\phi^2.$
But in terms of the variable $\phi^r$, 
the path integral represents the evolution of a
density matrix a la Schwinger-Keldysh,
and the path integral (\ref{EA-SH}) should be 
regarded as giving a probability, not an amplitude
for the configuration $\phi^r$. 
Based on this interpretation, we wrote it as $P$.

The classical limit with $T_H$ fixed corresponds to 
replacing $K(t,t')$ by $T_H \delta(t-t')$.
In this limit,
the probability for
 a trajectory $\Gamma_\tau[\phi^r]$ 
with an initial value  $\phi^r_{\text{ini}}$
to be realized is given by
\begin{align}
P[\Gamma_\tau|\phi^r_{\text{ini}}]= \exp \left[ -\frac{1}{4 T_H} \int_{\Gamma_\tau} dtd\Omega r_\epsilon^2 \left[
(\partial_t-\partial_{r^\ast})\phi^r(t) \right]^2
\right] \prod_{t, r>r_\epsilon, (l,m)} \delta\left[(\partial_t^2-\partial_{r_\ast}^2+V_l)\phi_{(l,m)}^r\right].
\end{align}
The ratio of the forward and the backward probabilities is now
given by
\begin{align}
\frac{P[\Gamma_\tau|\phi^r_{\text{ini}}]}{P[\Gamma^\ast_\tau|\phi^r_{\text{fin}}]}
= \exp \left[
\frac{1}{T_H} \int_{\Gamma_\tau}dtd\Omega r_\epsilon^2 \dot{\phi}^r(t) \partial_{r^\ast} \phi^r(t)
\right]
\end{align}
and reproduces the previous result (\ref{ratio_of_trajectory}).
The exponent is proportional to the energy flowing into the 
black hole across the horizon, and interpreted as an
entropy increase of the black hole.

More generally, if we do not replace $K(t,t')$ by the white noise,
the ratio becomes 
\begin{align}
\frac{P[\Gamma_\tau|\phi^r_{\text{ini}}]}{P[\Gamma^\ast_\tau|\phi^r_{\text{fin}}]} 
= \exp \left[ 
\int dt dt' d\Omega r_\epsilon^2 \dot{\phi}^r(t) K^{-1}(t,t') \partial_{r^\ast} \phi^r(t')
\right],
\end{align}
which is nonlocal in time.
By expanding the kernel in terms of derivatives of the delta functions,
the exponent receives corrections to the energy flow.
These corrections can be interpreted as flows of higher-spin currents
(operators containing higher derivatives of fields)
into the black hole.
These terms vanish after taking a long-time average, but remain
for a short time scale.
Applying the fluctuation theorems with the nonlocal modification
of the kernel, 
the entropy increase of the black hole receives higher derivative
corrections. A geometric interpretation of these corrections is 
interesting. 

Another important quantum correction is the violation of the 
Markovian assumption.
If the path integral is not dominated by classical paths,
we need to sum over all possible sequences of configurations
 at the level of amplitudes, 
instead of considering probabilities at the classical level.
We also need to generalize the fluctuation theorem themselves
at the fully quantum level. We hope to come back to these
issues in near future.

\subsection{Steady State Fluctuation Theorem in Black Holes}
So far, we have applied the fluctuation
theorem to  a scalar field in an
equilibrium distribution and disturbance around it.
In realizing such a  situation, we have put the black hole in a box
with an adiabatic (insulating) wall.
Instead we can consider a steady state with a constant 
(but very small) energy flow
from a black hole to outside.
This can be realized by putting a black hole in a box
in contact with a thermal bath
with a slightly lower (or higher) 
temperature than the Hawking temperature.
We set the temperature at the wall of the box as
$T_w (< T_H)$.
Since the energy flow is assumed to be very small,
we neglect  backreactions of the energy transfer
to the black hole itself.
The fluctuation theorem for the steady state is reviewed 
in the appendix \ref{A-SSFT}.

First we start from the following discretized form of the
equations of motion;
\begin{align}
\left\{
\begin{array}{l}
\gamma_0 \dot{x}_0=-k(x_0-x_1) -\sqrt{2}\xi_0 \ ,\  \langle\xi_0(t)\xi_0(t')\rangle=\gamma_0 T_H \delta (t-t')\\
\ddot{x}_1 =-k(x_1-x_2)+k(x_0-x_1) - V_l(r_1)x_1\\
\vdots  \\
\ddot{x}_N =-\gamma\dot{x}_N-k(x_N-x_{N+1})+k(x_{N-1}-x_N) 
- V_l(r_N)x_N -\sqrt{2}\xi \\
\hspace{20em} ,\langle\xi(t)\xi(t')\rangle=\gamma T_w \delta (t-t')\\
x_{N+1}\equiv 0. 
\end{array}
\right.
\end{align}

\begin{figure}[ht]
\begin{center}
\begin{overpic}[scale=0.6]{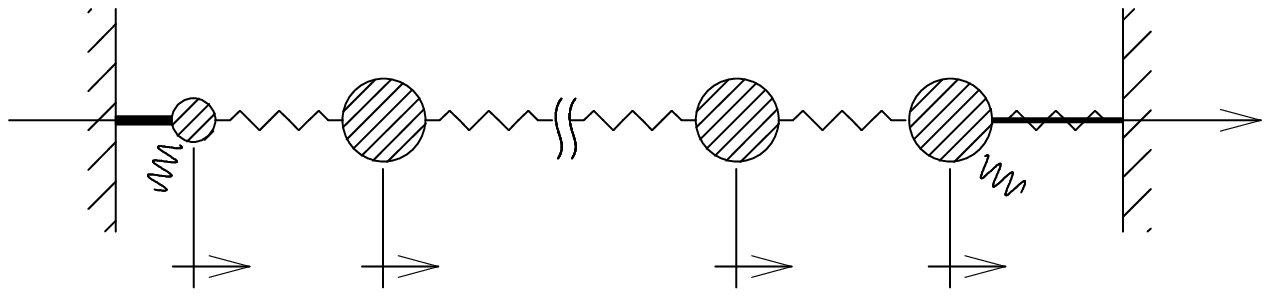}
\put(10,4){$\xi_0$}
\put(10,16.5){$\gamma_0$}
\put(21,16){$k$}
\put(38,16){$k$}
\put(49,16){$k$}
\put(65,16){$k$}
\put(82,16){$k$}
\put(16,4){$x_0$}
\put(32,4){$x_1$}
\put(59,4){$x_{N-1}$}
\put(75.5,3.5){$x_{N}$}
\put(81,6.5){$\xi$}
\put(84.5,9){$\gamma$}
\put(100,13){$r_\ast$}
\end{overpic}
\caption{A discretized model of a scalar field in a box. The wall at 
$r=r_B$ is in contact with a thermal bath with temperature 
$T_w$ slightly 
lower than the Hawking temperature. Then there is an energy flow
from the black hole to the outer thermal bath.}
\end{center}
\label{modelB}
\end{figure} 
The only difference from the previous model is the equation 
for the variable $x_N$ at the wall. 
The variable in this model interacts with the thermal bath
at temperature $T_w$. A redundant variable $x_{N+1}$
is introduced for simplifying the above equation.
The corresponding Fokker-Planck equation is given by
\begin{align}
\partial_t P&=\partial_{x_0}\left[\frac{1}{\gamma_0} \frac{\partial U}{\partial x_0}P+\frac{T_H}{\gamma_0}\partial_{x_0}P \right]
+\sum_{i=1}^{N-1}\left[\partial_{x_i}(-v_i P)+\partial_{v_i}\left( \frac{\partial U}{\partial x_i}P\right)\right]\nn
&\hspace{5em}+\partial_{x_N}(-v_N P)+\partial_{v_N}\left(\gamma 
v_N+ \frac{\partial U}{\partial x_N}P
+\gamma T_w\partial_{v_N}P \right).
\end{align}
The solution can be written in terms of the following path integral
\begin{align}
P(x_{\text{fin}} ,\tau| x_{\text{ini}}, 0)&=
\int_{x_{\text{ini}}}^{x_{\text{fin}}}{\cal D}\Gamma_\tau
 e^{-\frac{1}{4\gamma_0 T_0}\int_{\Gamma_\tau} dt\left(\gamma_0\dot{x}_0+\partial_{x_0}U\right)^2}
\prod_{t}\prod_{i=1}^{N-1}\delta\left(\ddot{x}_i+\partial_{x_i}U\right)\nn
&\hspace{5em}\times
e^{-\frac{1}{4\gamma T_w}\int_{\Gamma_\tau} dt\left(\ddot{x}_N+\gamma\dot{x}_N+\partial_{x_N}U\right)^2}|_{x_{N+1}=0} .
\end{align}
The ratio of the probabilities for a single partial wave with
$(l,m)$
is given by 
\begin{align}
\frac{P[\Gamma_\tau|x_i]}{P[\Gamma_\tau^\ast|x_f]}
&=\exp\left[-\frac{1}{T_H}\int_{\Gamma_\tau} dt \dot{x}_0\partial_{x_0}U 
-\frac{1}{T_w}\int_{\Gamma_\tau} dt\dot{x}_N(\ddot{x}_N+\partial_{x_N}U) \right]_{|x_{N+1}=0} .
\label{Key2}
\end{align}
In addition to the energy flow at the horizon, there is another contribution
from the wall. 
The potential $U(x)$ is written as a sum of three terms;
\begin{align}
U(x) = U_1(x_0,x_1,\cdots , x_{N-1})+U_{12}(x_{N-1}, x_N)+U_2(x_N ,x_{N+1})
\end{align}
where
\begin{align}
U_1(x_0,x_1,\cdots , x_{N-1})&=\frac{1}{2}\sum_{i=1}^{N-1}
\left[\left( \frac{x_i-x_{i-1}}{d}\right)^2+V_l(r_i)x_i^2 \right], \nn
U_{12}(x_{N-1}, x_N)&=
\frac{1}{2}\left( \frac{x_N-x_{N-1}}{d}\right)^2, \nn 
U_2(x_N, x_{N+1})&= \frac{1}{2}\left[\left( \frac{x_{N+1}-x_{N}}{d}\right)^2+V_l(r_N)x_N^2 \right].
\end{align}

We turn on the potential 
$U_{12}$ at the wall during a time interval between
$t=0$ and $t=\tau.$
This can be realized by introducing the external parameter
controlling the potential $U_{12}$
such as 
\begin{align}
U_{12}(x_{N-1}, x_N; \lambda^F_t)&=
\theta\left(\tfrac{\tau_-}{2}-|t-\tfrac{\tau}{2}|\right)U_{12}(x_{N-1}, x_N).
\end{align}
Then the variables $(x_0, x_1, \cdots , x_{N-1})$ are decoupled 
from $x_N$ when $t<0$ and $t>\tau$.
Since the external thermal bath is decoupled for a long time 
during $t<0$,
the state can be considered in the equilibrium at $t=0$.
The ratio of the probabilities of the initial distributions is, hence,
given by 
\begin{align}
\frac{P^{\text{eq}}(x_{\text{ini}}) }{P^{\text{eq}}(x_{\text{fin}})}
&=\exp\Biggr[
-\frac{1}{T_H}\left(\frac{1}{2}\sum_{i=1}^{N-1}(\dot{x}_{i, \text{ini}}^2
-\dot{x}_{i, \text{fin}}^2) +U_1(x_{\text{ini}})-U_1(x_{\text{fin}})\right)\nn
&\hspace{5em}-\frac{1}{T_w}\left(\frac{1}{2}(\dot{x}_{N, \text{ini}}^2
-\dot{x}_{N, \text{fin}}^2) +U_2(x_{\text{ini}})-U_2(x_{\text{fin}})\right)
\Biggr].
\label{steadyinitial}
\end{align}
The second terms are canceled by 
the following terms in eq.(\ref{Key2}) 
\begin{align}
\int_{\Gamma_\tau} dt\dot{x}_N(\ddot{x}_N+\partial_{x_N}U_2(x_N))
&=\left[\frac{1}{2}m\dot{x}_N^2+U_2(x_N)\right]_{\text{ini}}^{\text{fin}}.
\end{align}
The remaining terms in (\ref{steadyinitial})
is, of course, independent of the duration $\tau$,
and can be neglected compared to other terms in (\ref{Key2}) 
that are proportional to $\tau$.

As a result, if take the leading contributions in the large $\tau$ limit
and neglect ${\cal O}(\tau^0)$ terms in the exponent,
we obtain 
\begin{align}
\frac{P[\Gamma_\tau|x_{\text{ini}}]P^{\text{eq}}(x_{\text{ini}})}{P[\Gamma_\tau^\ast|x_{\text{fin}}]P^{\text{eq}}(x_{\text{fin}})}
&=\exp\left[-\frac{1}{T_H}\int_{\Gamma_\tau} dt \dot{x}_0\partial_{x_0}U_1 
-\frac{1}{T_w}\int_{\Gamma_\tau} dt\dot{x}_N \partial_{x_N}U_{12}  \right]\nn
&=\exp\left[-\frac{1}{T_H}\int_{\Gamma_\tau} dt \dot{x}_0kd \Delta^- x_0
-\frac{1}{T_w}\int_{\Gamma_\tau} dt\dot{x}_N kd \Delta^-x_N \right].
\end{align}
Because of the energy conservation for a steady state,
we have the relation $\int dt\dot{x}_N \Delta^-x_N =-\int dt \dot{x}_0\Delta^- x_0$. 
In the contimuum limit $N\to \infty$ with the scalings explained before,
the logarithm of the above ratio  becomes
\begin{align}
&-\frac{1}{T_H}\int_{\Gamma_\tau} dt \dot{x}_0kd \Delta^- x_0
-\frac{1}{T_w}\int_{\Gamma_\tau} dt\dot{x}_N kd \Delta^-x_N \nn
&=
(\beta_w-\beta_H)\int_{\Gamma_\tau} dt \partial_t\phi^r_{(l,m)}(t,r_\epsilon)\partial_{r_\ast}\phi^r_{(l,m)}(t,r_\epsilon)
\equiv \Delta\beta \tau \bar{J}_{(l,m)}[\Gamma_\tau].
\end{align}
We have defined $\Delta \beta\equiv \beta_w-\beta_H$, which is positive 
from our assumption $T_H> T_w$.
By summing all the contributions from the partial waves with $(l,m)$,
we have
\begin{align}
\bar{J}[\Gamma_\tau]\equiv
\frac{1}{\tau} \int_{\Gamma_\tau} dt d\Omega \  r_\epsilon^2\partial_t\phi^r(t,r_\epsilon,\Omega)\partial_{r_\ast}\phi^r(t,r_\epsilon,\Omega)=
\frac{1}{\tau}\int_{\Gamma_\tau} dt d\Omega \ r_\epsilon^2 T^r_t(t, r_\epsilon,\Omega),
\end{align}
where we have used the definition of the 
energy momentum tensor $T^r_t=\partial_t\phi^r \partial^r\phi^r=\partial_t\phi^r \partial_{r_\ast}\phi^r$.
$\bar{J}[\Gamma_\tau]$ is a current flowing at the horizon out of the black hole. 
From the setting $T_H>T_w$, the averaged current is positive, but
it can take  both positive or negative values
because of fluctuations of absorption and emission by the Hawking radiation.

We now have established the steady state fluctuation theorem 
in the black hole background as
\begin{align}
\lim_{\tau\to\infty}\frac{1}{\tau}\left[\frac{\rho(\bar{J}_\tau , \Delta\beta)}{\rho(-\bar{J}_\tau , \Delta\beta)}\right]
=\Delta \beta \bar{J}_\infty.
\end{align}

For the definitions of $\rho$, 
see eq. \ref{SSFT1}. 
The theorem can be restated 
in terms of a generating function $Z(\alpha_\tau , \Delta \beta)$,
and leads to various 
relations between the response coefficients $L^{(1)} ,L^{(2)},\cdots  $ 
defined by $\langle \bar{J}_\infty \rangle=L^{(1)}\Delta \beta +L^{(2)}/2 (\Delta \beta)^2+\cdots $
and correlator of currents
$\langle J(t)J(t') \rangle$ .
For more details, see the appendix \ref{A-SSFT}.

In our case, these relations  lead to the following relations;
\begin{align}
L^{(1)}&= \frac{1}{2}\int_0^\infty dt \int d\Omega \ r_\epsilon^4
 \langle T^r_t(t, r_\epsilon) T^r_t(0, r_\epsilon)\rangle_{|\Delta \beta=0}\\
L^{(2)}&= \lim_{\tau\to\infty}\frac{1}{2\tau}\int^\tau dtdt' 
\int d\Omega \ r_\epsilon^4
 \partial_{\Delta\beta}\langle T^r_t(t, r_\epsilon) T^r_t(t', r_\epsilon)
\rangle_{|\Delta \beta=0}\\
&\vdots \nonumber
\end{align}
The first relation 
is the Green-Kubo relation for the energy current flowing 
at the horizon $r=r_\epsilon$.
The second one is a non-linear generalization, and the evaluation of the
right hand side needs the derivative of the correlation function with
respect to the temperature difference.
This means that the information of the equilibrium distribution 
at $\Delta \beta =0$
is not sufficient to obtain the non-linear response function
of the current.

\section{Summary}
In this paper, we  derived a stochastic equation 
with a dissipative term and a noise
for a scalar field in a black hole background.
The dissipation comes from the ingoing boundary condition
at the horizon while the noise comes from the Hawking radiation.
The stochastic equation can be derived by considering a stretched
horizon and integrating variables between the horizon and the 
stretched horizon.
The stochastic equation describes the dynamics of the scalar
field in the limit $\hbar \to 0$ with
the Hawking temperature $T_H=\hbar \kappa/2\pi$ kept finite.
We then applied  the non-equilibrium fluctuation theorems,
developed in the statistical physics,  
to the above stochastic equation in the black hole background.
We consider two cases. One is a scalar field confined in a box
 with an insulating wall. The system is relaxed to an equilibrium 
 state at the Hawking temperature.
The fluctuation theorem leads to the second law of the black hole 
thermodynamics after taking a thermal average.
The other case is a scalar field in a box in contact with
a heat bath with a different temperature from $T_H$.
Then there is an energy flow between the horizon and the outer boundary.
The fluctuation theorem leads to the Green-Kubo relation and its 
nonlinear generalizations.

We have used an approximation of
 replacing a nonlocal (colored) noise correlation by a white noise.
 We furthermore approximated the dynamical evolution 
of the scalar field by a classical  Markovian process.
These approximations are valid in the classical limit $\hbar \to 0$
with the Hawking temperature $T_H$ fixed.
In this sense, quantum effect is partially taken into account
through the Hawking radiation.
The  results such as the ordinary second
law of the black hole thermodynamics or the Green-Kubo relation
are derived only in such approximations.
As mentioned in Sec \ref{Sec-Memory}, the nonlocal noise correlation
 leads to a deviation of the black hole entropy appearing in the 
 second law of thermodynamics.
We hope to come back to these issues in near future.

Finally it is interesting to generalize our results for
a scalar field to a vector or a gravitational field
and obtain  quantum corrections to the membrane action \cite{Thorne:1986, Parikh:1997ma}.
The absorption of energy across the stretched horizon is 
known to give  dissipative equations such as the Ohm's law
or Navier-Stokes equation on the membrane. 
If the Hawking radiation is included, these equations 
must receive quantum corrections as noise terms.
Then the Hawking radiation may be
interpreted as the anomaly inflow of the 
 gravitational and gauge anomalies. 
In this sense, 
the quantum membrane action  will be analogous to 
the edge state of quantum Hall effect. 
It is furthermore interesting if we can relate such 
a quantum membrane interpretation to the black hole entropy.

\vskip6mm
\noindent
{\bf Acknowledgments}\\
We would like to thank S. Zhang for  collaboration at the early stage of the work.
We thank the participants of the workshop "Towards New Developments in Field and String Theories" 
at RIKEN in December 17-19, 2010 for valuable discussions. 
We thank our colleagues, especially Y. Kitazawa and H. Kitamoto, for discussions, 
M. Hotta and M. Morikawa for inviting us to a wonderful workshop
at Kusatsu in March 6-10, 2011.
We furthermore acknowledge useful discussions with
experts in the statistical physics, T. Sagawa, S. Sasa
and H. Tasaki. 
The research by S.I. is supported in part by the Grant-in-Aid for Scientific Research (19540316) from MEXT, Japan.
We are also supported in part by "the Center for the Promotion of Integrated Sciences (CPIS) "  of Sokendai.

\appendix
\section{Path integral form of the Fokker-Planck equation \label{app-OM}}
In this appendix we derive the path integral form (\ref{Path})
of the solution
 to the Fokker-Planck equation;
\begin{align}
\partial_t P(x,v,t|x_0,v_0,0) &= \hat{L}_{FT} P(x,v,t|x_0,v_0,0)  \nonumber \\
 &= \partial_x\left(-v P\right)
 +\partial_v\left[\left(\frac{\gamma}{m}v+\frac{1}{m}\frac{\partial V}{\partial x} \right)P \right]
+\partial_v^2\left(\frac{\gamma T}{m^2}P\right) .
\end{align}
For a small time-interval, it can be written as
\begin{align}
&P(x,v,\Delta t|x_0,v_0, 0)=e^{\Delta t \hat{L}_{FP}}\delta(x-x_0)\delta(v-v_0)\nn
&\sim \int\frac{dk_x dk_v}{(2\pi)^2}\left[1+\Delta t \left[-v_0ik_x+\left(\tfrac{\gamma}{m} v_0 +\tfrac{1}{m}\tfrac{\partial V(x_0)}{\partial x}\right)ik_v
-\tfrac{\gamma T}{m^2}k_v^2 \right] \right]e^{ik_x(x-x_0)+ik_v(v-v_0)}\nn
&\sim \int\frac{dk_x dk_v}{(2\pi)^2} \exp\left[i\Delta t k_x\left(\tfrac{x-x_0}{\Delta t} -v_0 \right)
-\Delta t \tfrac{\gamma T}{m^2}\left(k_v-i\tfrac{m}{2\gamma T}\left(m\tfrac{v-v_0}{\Delta t}
+\gamma v_0 +\tfrac{\partial V(x_0)}{\partial x}\right)\right)^2\right]\nn
&\hspace{6em} \times\exp\left[-\tfrac{\Delta t}{4\gamma T}
\left(m\tfrac{v-v_0}{\Delta t}+\gamma v_0 +\tfrac{\partial V(x_0)}{\partial x}\right)^2 \right]\nn
&=\sqrt{\tfrac{2\pi m^2}{\Delta t\gamma T}}\delta(\dot{x}_0-v_0)
\exp\left[-\tfrac{\Delta t}{4\gamma T}
\left(m\dot{v}_0+\gamma v_0 +\tfrac{\partial V(x_0)}{\partial x}\right)^2\right]
\end{align}
Then by using the Chapman-Kolmogorov equation 
$P(X_3|X_1)=\int dX_2 P(X_3|X_2)P(X_2|X_1)$ which is
equivalent to an insertion of the complete set
and integrating over $v$,
we obtain the path integral form as follows;
\begin{align}
P(x,t|x_0,0)=\int_{x(0)=x_0}^{x(t)=x}{\cal D}x \exp \left[-\tfrac{1}{4\gamma T}\int_0^t dt'
\left(m\ddot{x}+\gamma \dot{x} +\tfrac{\partial V}{\partial x}\right)^2\right].
\end{align}
If we use the Langevin equation (\ref{Langevin}),
the path integral is equivalent to the noise average  with the weight function
in eq. (\ref{NPI}).

\section{Noise correlation and Hawking radiation \label{app-HR}}
The noise correlation induced in the effective equation of motion 
for the boundary field at the stretched horizon
$r=r_H+\epsilon$ can be interpreted 
as the Hawking radiation. Here we first review the method to determine
the energy-momentum tensor in the black hole background by using
the trace anomaly of the energy-momentum tensor and the
regularity condition at the horizon \cite{PhysRevD.15.2088}, and then generalize the method
to determine higher spin currents \cite{Iso:2007kt, Iso:2007hd, Iso:2007nc, Iso:2007nf}. 

In two dimensions, the  trace of the energy-momentum tensor
of a single scalar field (i.e. the central charge is $c=1$)
has an anomaly term proportion to the scalar curvature $R$
\begin{align}
T^\mu_\mu&=\frac{1}{24\pi}R.
\end{align}
Writing the metric in the conformal gauge
 $ds^2=e^{\varphi (u,v)}(-dudv)$, 
the equation becomes
 $T_{uv}=-\frac{1}{24\pi}\partial_u\partial_v \varphi$.
By combining with the
 conservation of the energy-momentum tensor $\nabla_\mu T^\mu_\nu =0$,
derivatives of the EM tensor
$\partial_vT_{uu}(u,v)$ and $\partial_uT_{vv} (u,v)$ 
can be written as follows;
\begin{align}
\partial_v T_{uu}&=\frac{1}{24\pi}\left[\partial_u^2\partial_v\varphi -(\partial_u\varphi)(\partial_u\partial_v\varphi)\right]\\
\partial_u T_{vv}&=\frac{1}{24\pi}\left[\partial_v^2\partial_u\varphi -(\partial_v\varphi)(\partial_u\partial_v\varphi)\right].
\end{align}
From these equations, we 
 can define a (anti-) holomorphic quantity
\begin{align}
t_{uu}(u)&\equiv T_{uu}-\frac{1}{24\pi}\left[\partial_u^2\varphi -\frac{1}{2}(\partial_u\varphi)^2\right]
\label{HEMT}
\\
t_{vv}(v)&\equiv T_{vv}-\frac{1}{24\pi}\left[\partial_v^2\varphi -\frac{1}{2}(\partial_v\varphi)^2\right].
\end{align}
They are often called (anti-) holomorphic energy-momentum tensors,
but their transformation laws are anomalous and not tensors
in the exact sense. 
Actually,
under a coordinate transformation from $(u,v)$ to $(U=U(u),V=V(v))$,
they transform as 
\begin{align}
 &t_{UU}(U)=\frac{1}{\left(\kappa U\right)^2}\left[t_{uu}(u)
+ \frac{1}{24\pi}\{U,u \}
\right],
\end{align} 
where $\{ U,u\}$ is the Schwarzian derivative,
\begin{align}
\{U, u\}\equiv \frac{\partial_u^3U}{\partial_uU}-\frac{3}{2}\left(\frac{\partial_u^2U}{\partial_uU}\right)^2.
\end{align}
In particularly, for the transformation from the Schwarzschild 
coordinates to the Kruskal ones, namely from
 $(u,v)$ to $(U,V)=(-\kappa^{-1}e^{-\kappa u} , \kappa^{-1}e^{\kappa v} )$,
the Schwarzian derivative becomes $\{U, u\}=-\kappa^2/2$.

Now, we impose the 
regularity condition at the horizon.
The energy momentum tensor $T_{UU}$ 
must behave regularly near the future horizon $U=0$
in the regular coordinates,
and so is $t_{UU}(U)$ since they are related regularly as (\ref{HEMT}).
The regularity condition, hence, imposes that $t_{uu}$ must behave as
\begin{align}
t_{uu}(u\to \infty)=\frac{\kappa^2}{48\pi}.
\end{align}
If we neglect the effect of scatterings of the 
outgoing fluxes (namely in the absence of the gray body factor), 
we can extrapolate the above flux at the
horizon to
the outgoing flux at $r\to\infty$
as 
\begin{align}
T_{uu}(r\to\infty)&=\frac{\kappa^2}{48\pi} =\frac{\pi}{12}T_H^2.
\end{align}
It is interpreted as the flux from the black body with the Hawking
temperature $T_H$,
\begin{align}
\int_0^\infty \frac{d\omega}{2\pi}\frac{\omega}{e^{\beta \omega}-1}
&=\frac{\pi}{12}T_H^2.
\label{EMflux}
\end{align}

The  transformation property of the holomorphic energy-momentum 
tensor can be also derived by considering the following 
point-splitting regularization,
\begin{align}
:t_{uu}(u): &\equiv \lim_{\delta\to 0}\left[\partial_u\phi (u+\tfrac{\delta}{2})\partial_u \phi(u-\tfrac{\delta}{2}) 
-\langle\partial_u\phi (u+\tfrac{\delta}{2})\partial_u \phi(u-\tfrac{\delta}{2}) \rangle \right]\nn
&=\lim_{\delta\to 0}\left[\partial_u\phi (u+\tfrac{\delta}{2})\partial_u \phi(u-\tfrac{\delta}{2}) 
+\frac{1}{4\pi \delta^2}\right],
\end{align}
where we have used the explicit form of the free boson propagator 
$\langle \phi(u) \phi(u')\rangle=- \ln(u-u')/4\pi$.
From this definition, we can relate it to the energy momentum 
tensor regularlized in the Kruskal ($U$) coordinate;
\begin{align}
&:t_{uu}(u): \nn
&=\lim_{\delta\to 0}\left[\partial_u U(u+\tfrac{\delta}{2})\partial_u U(u-\tfrac{\delta}{2})
 \partial_U\phi (U(u+\tfrac{\delta}{2}))\partial_U \phi(U(u-\tfrac{\delta}{2})) 
+\frac{1}{4\pi \delta^2}\right]\nn
&=\lim_{\delta\to 0}\left[\partial_u U(u+\tfrac{\delta}{2})\partial_u U(u-\tfrac{\delta}{2})
 \left( t_{UU}(U)-\frac{1}{4\pi}\frac{1}{(U(u+\tfrac{\delta}{2})-U(u-\tfrac{\delta}{2}) )^2}\right)
+\frac{1}{4\pi \delta^2}\right]\nn
&=(\partial_u U)^2 :t_{UU}(U):_K-\frac{1}{24\pi}\{U , u\}.
\end{align}
Namely, the Schwarzian derivative is nothing but the difference 
of the normal orderings in different coordinates.

The energy flux (which corresponds to the first moment of the 
thermal spectrum (\ref{EMflux})) can be generalized to a flux of 
a higher spin current with a higher moment, and its generating
function can be defined as a correlation function 
of the scalar field;
\begin{align}
J(u , u+a)&\equiv \sum_{n=0}^\infty\frac{a^n}{n!} :\partial_u\phi(u)\partial^{n+1}\phi(u) :\nn
&=:\partial_u\phi(u) \partial_u\phi(u+a): \ .
\label{J}
\end{align}
The normal ordering $: \  : $ is defined similarly to $t_{uu}(u)$ by 
\begin{align}
:\partial_u\phi(u)\partial_u\phi(u):
\ \equiv\lim_{u'\to u}\left[
\partial_u\phi(u)\partial_u\phi(u')-\langle \partial_u\phi(u)\partial_u\phi(u')\rangle
 \right].
\end{align}
Then we can show that 
 $J(u, u+a)$ 
 transforms under  the coordinate transformation from $u$ to $U(u)$ as
\begin{align}
J(u,u+a)&=
\partial_uU(u) \partial_uU(u+a) J(U(u), U(u+a))
+\frac{1}{4\pi}\left[-\frac{\kappa^2}{4 \sinh^2\frac{\kappa a}{2}} +\frac{1}{a^2}\right].
\end{align}
Similarly to the energy flux discussed before,
the regularity condition at the future horizon
fixes  the value of $J(u,u+a)$ at $U=0$ as
\begin{align}
J(u,u+a)|_{r=r_H}=:\partial_u\phi(u) \partial_u\phi(u+a): =
\frac{1}{4\pi}\left[-\frac{\kappa^2}{4 \sinh^2\frac{\kappa a}{2}} +\frac{1}{a^2}\right].
\label{value_of_J}
\end{align}
This can be interpreted as a correlation function of 
$ \partial_u \phi(u) $
and $\partial_u \phi(u+a)$ on the Kruskal vacuum.

In Section \ref{sec-Lan}, we have shown that the scalar field
obeys a stochastic equation of motion 
\begin{align}
 \partial_u \phi(t-r^\ast)_{|r=r_H+\epsilon}=-
\sqrt{2}\xi(t).
\end{align} 
at the stretched horizon.
Since the equation is independent of the value of 
$\epsilon$, we can safely take 
 $\epsilon \rightarrow 0$ limit. 
Then the value of the generating function $J(u,u+a)$
for the higher spin fluxes discussed above
is equivalent to the noise correlation 
$2 \langle \xi(t)\xi(t+a)\rangle$ of the Langevin equation
at the horizon.
The functional forms 
 are equal, though the coefficients
 are different by a factor 4.
Reason for the discrepancy is now under study.

\section{The Steady State Fluctuation Theorem \label{A-SSFT}} 
In this appendix, we consider the fluctuation theorem for a steady state and
derive the Green-Kubo formula. 

Assume that  we have two variable $x_1 , x_2$,
and each of them is
in contact with a different thermal bath with temperature $T_1$ and $T_2$.
We further assume that the dynamics is governed by the set of 
Langevin equations such as
\begin{align}
m_1\dot{v}_1+\frac{\partial V}{\partial x_1}+\gamma_1 v_1= \xi_1
\ \ \  , \ \ \langle\xi_1(t)\xi_1(t') \rangle=2\gamma_1 T_1 \delta(t-t')
\nn
m_2\dot{v}_2+\frac{\partial V}{\partial x_2}+\gamma_2 v_2= \xi_2
\ \ , \ \ 
\langle\xi_2(t)\xi_2(t') \rangle=2\gamma_2 T_2 \delta(t-t').
\end{align}
Here, $V(x_1, x_2; \lambda_t^F)$ is an interaction potential between the 
two variables.
The corresponding Fokker-Planck equation of the system can be obtained straightforwardly.
The trajectory $\Gamma_\tau$ is also generalized as 
$\Gamma_\tau=\{x(t)=(x_1(t), x_2(t))|
x(0)=(x_1(0), x_2(0))=(x^1_{\text{ini}}, x^2_{\text{ini}})\}$.
Then the solution to the Fokker-Planck equation gives 
probabilities of the forward and the reversed protocols, and 
the ratio is given by
\begin{align}
\frac{P^F[\Gamma_\tau|x_{\text{ini}}]}{P^R[\Gamma_\tau^\ast|x_{\text{fin}}]}
&=\exp\left[-\frac{1}{T_1}\int_{\Gamma_\tau}dt \dot{x}_1\left(m_1\ddot{x}_1+\frac{\partial V(x;\lambda^F_t)}{\partial x_1}\right)
-\frac{1}{T_2}\int_{\Gamma_\tau}dt \dot{x}_2\left(m_2\ddot{x}_2+\frac{\partial V(x;\lambda^F_t)}{\partial x_2}\right)
\right].
\end{align}
We have assumed that the two variables are decoupled before $t=0$ and after $t=\tau$;
the interaction potential $V$ vanishes  at $t<0$ and $t>\tau$.
The initial distribution of the total system is given by a product of 
the equilibrium distributions of each system 
$P^{\text{eq}}(x_{\text{ini}})=P^{\text{eq}}(x^1_{\text{ini}})P^{\text{eq}}(x^2_{\text{ini}})$.
The forward protocol is expressed as
\begin{align}
V(x;\lambda^F_t)&=V_1(x_1)+V_2(x_2)+f(\lambda^F_t)V_{12}(x_1-x_2)
\end{align}
where
\begin{align}
f(\lambda^F_t)&=\theta\left(\frac{\tau_-}{2}-|\lambda^F_t-\frac{\tau}{2}|\right), \ \  \lambda^F_t=t.
\end{align}
$\tau_-$ means $\tau-\epsilon $ for $ 0<\epsilon \ll \tau$.
Function $f(t)$ satisfies $f(t=0)=f(t=\tau)=0$  and $f(0< |t|< \tau)=1$,
so that
the interaction  switches on at $t=0$ and  off at $t=\tau$.
This protocol has the reversal symmetry $f(\lambda^F_t)=f(\lambda^F_{\tau-t})$.

In considering the large  interval limit $\tau \to \infty$, 
the energy transfer such as $\int dt \dot{x}_1 \partial_{x_1} V_{12}(x_1-x_2)$ 
(or $\int dt \dot{x}_2 \partial_{x_2} V_{12}(x_1-x_2)$) grows linearly in $\tau$. 
On the other hand $\Delta E_1=\int dt \dot{x}_1(m_1\ddot{x}_1+\partial_{x_1} V_1(x_1) )
=
(\frac{1}{2}m_1\dot{x}_1^2+V_1(x_1) )_{t=\tau} - (\frac{1}{2}m_1\dot{x}_1^2+V_1(x_1) )_{t=0} $ 
or $\Delta E_2= \int dt \dot{x}_2 (m_2\ddot{x}_2+\partial_{x_2} V_2(x_2) )$
is at most ${\cal O}(\tau^0)$. 
If each system becomes stationary after taking $\tau \rightarrow \infty$, 
the change in the energy  of each system  vanishes.
Hence we can drop the term 
 $P^{\text{eq}}(x_{\text{ini}})/P^{\text{eq}}(x_{\text{fin}})$
and a contribution of $\Delta E_i$  in $P[\Gamma_\tau|x_{\text{ini}}]/P[\Gamma^\ast_\tau|x_{\text{fin}}]$
when we evaluate the quantity
\begin{align}
\lim_{\tau\to \infty}\frac{1}{\tau}\ln \left(
\frac{P[\Gamma_\tau|x_{\text{ini}}] P^{\text{eq}}(x_{\text{ini}})}{P[\Gamma_\tau^\ast|x_{\text{fin}}] P^{\text{eq}}(x_{\text{fin}})} 
 \right).
\end{align}
In addition, 
we have $\int_{\Gamma_\tau}dt \dot{x}_1\partial_1 V_{12} \sim
 -\int_{\Gamma_\tau}dt \dot{x}_2\partial_2 V_{12} +  {\cal O}(\tau^0)$. 
Therefore
we can write the ratio of the probabilities only in terms of the 
energy current
 defined by  $\bar{J} [\Gamma_\tau] \equiv \frac{1}{\tau}\int_{\Gamma_\tau}dt \dot{x}_1\partial_1 V_{12}$.
Writing the temperature difference as $\Delta \beta\equiv \beta_2-\beta_1$,
we obtain the following relation; 
\begin{align}
\rho(\bar{J}_\tau, \Delta \beta)&\equiv
\int{\cal D}x P[\Gamma_\tau|x_{\text{ini}}] P^{\text{eq}}(x_{\text{ini}})\delta(\bar{J}_\tau-\bar{J}[\Gamma_\tau])\nn
&\simeq \int{\cal D}x P[\Gamma^\ast_\tau|x_{\text{fin}}] P^{\text{eq}}(x_{\text{fin}}) e^{\tau \Delta\beta\bar{J}[\Gamma_\tau]}
\delta(\bar{J}_\tau-\bar{J}[\Gamma_\tau])\nn
&=e^{\tau \Delta\beta\bar{J}_\tau}
\int{\cal D}x P[\Gamma^\ast_\tau|x_{\text{fin}}] P^{\text{eq}}(x_{\text{fin}}) \delta(\bar{J}_\tau+\bar{J}[\Gamma^\ast_\tau])\nn
&=e^{\tau \Delta\beta\bar{J}_\tau}\rho(-\bar{J}_\tau, \Delta \beta).
\label{SSFT1}
\end{align}
The steady state fluctuation theorem can be written  as
\begin{align}
\lim_{\tau\to \infty}\frac{1}{\tau}\ln\left[
\frac{\rho(\bar{J}_\tau, \Delta \beta)}{\rho(-\bar{J}_\tau , \Delta \beta)}\right]
&=\Delta\beta \bar{J}_\infty.
\label{STFT}
\end{align}

From this relation,
we can derive  the Green-Kubo relation and its non-linear generalizations.
By using  the generating function 
\begin{align}
Z(\alpha_\tau , \Delta \beta)\equiv \ln\left(
\int_{-\infty}^\infty d\bar{J}_\tau e^{i\tau \bar{J}_\tau \alpha_\tau} \rho(\bar{J}_\tau, \Delta \beta)
\right),
\end{align}
the steady state fluctuation theorem (\ref{SSFT1}) can be recast into
\begin{align}
Z(\alpha_\tau +i\Delta \beta , \Delta \beta)=Z(-\alpha_\tau , \Delta \beta).
\label{SSFT}
\end{align}
Taking a derivative of  both sides
with respect to  $\Delta \beta$ and setting $\Delta \beta=0$, we have
\begin{align}
\partial_{\Delta \beta}\left[Z(\alpha_\tau , 0)-Z(-\alpha_\tau , 0) \right]
&=-i\partial_{\alpha_\tau}Z(\alpha_\tau , 0).
\label{Green}
\end{align}

The generating function can be expanded in terms of the 
 correlators of $\bar{J}_\tau$ as 
\begin{align}
Z(\alpha_\tau , \Delta\beta)=\sum_{n=1}^\infty\frac{(i\tau \alpha_\tau)^n}{n!}G_n(\Delta\beta).
\end{align}
$G_n(\beta)$ gives a connected Green function of the averaged current
\begin{align}
 \bar{J}_\tau = \frac{1}{\tau} \int_0^\tau dt J(t).
\end{align}
Now the equation (\ref{Green}) is rewritten in the following form;
\begin{align}
\left[1-(-1)^n\right]\partial_{\Delta\beta}G_n(0)=\tau G_{n+1}(0).
\end{align}
We further expand the one-point function of $\bar{J}_\tau$,
which gives an expectation value of the current, with respect to 
the inverse temperature difference $\Delta \beta$ as
\begin{align}
G_1(\Delta \beta)\equiv \sum_{m=0}^\infty \frac{L^{(m)}}{m!}(\Delta \beta)^m.
\end{align}
For $n=0$, we have a trivial identity $G_{1}(0)=L^{(0)}=0$.
For $n=1$, the Green-Kubo relation is derived;
\begin{align}
L^{(1)}&=
\frac{1}{2 \tau}\int_0^\tau dt dt' \langle J(t) J(t')\rangle_{|\Delta\beta=0}\nn
&\xrightarrow{\tau\to\infty} \frac{1}{2}\int_0^\infty dt \langle J(t) J(0)\rangle_{|\Delta\beta=0} .
\end{align}
When $\Delta\beta=0$, 
the system is described by the equilibrium distribution function
$P^{\text{eq}}(x)=e^{-\beta E_{\text{tot}}(x)}/Z , \beta=\beta_1=\beta_2$
and an expectation value of a function $F(x(t))$ is given by
 $\langle F(x(t)) \rangle_{|\Delta\beta=0}=\int{\cal D}x P^{\text{eq}}(x(t)) F(x(t))$.
In the large $\tau$ limit, the correlator  $\langle J(t)J(t')\rangle_{|\Delta\beta=0}$ depends only on $(t-t')$.

We can also obtain the expression of $L^{(2)} , L^{(3)}, \cdots$ by taking
further derivatives of the equation (\ref{SSFT})
with respect to $\Delta \beta$. 
For instance, we can derive 
\begin{align}
\partial_{\Delta \beta}^2\left[Z(\alpha_\tau , 0)-Z(-\alpha_\tau , 0)\right]
&=-i\partial_{\alpha_\tau}\partial_{\Delta\beta}\left[
Z(\alpha_\tau , 0)+Z(-\alpha_\tau , 0)\right]\nn
\Rightarrow \left(1-(-1)^n\right)\partial_{\Delta \beta}^2K_n(0)
&=\tau \left(1+(-1)^{n+1}\right)\partial_{\Delta \beta}K_{n+1}(0).
\end{align}
For $n=1$, we have
\begin{align}
L^{(2)}&=\lim_{\tau\to \infty}
\frac{1}{2\tau}\int_0^\tau dt dt' \partial_{\Delta \beta}\langle J(t) J(t')\rangle_{|\Delta\beta=0}.
\end{align}
These nonlinear generalizations can be systematically obtained by using the steady state
fluctuation theorems.  
We apply these expansion method to a system of a black hole and matter
 to obtain the Green-Kubo relation for a thermal current in the body of the paper.

\bibliography{BHFT}

\end{document}